\documentclass[12pt]{article}
\DeclareMathAlphabet{\scr}{U}{rsfs}{m}{n}
\usepackage{latexsym}
\usepackage[mathscr]{eucal}
\usepackage{amsfonts}
\usepackage{amscd}
\usepackage{cite}
\usepackage{array}   
\usepackage{amssymb}
\usepackage{colordvi}
\usepackage[centertags]{amsmath}
\usepackage{enumerate}
\usepackage{graphicx}
\usepackage{booktabs}
\usepackage{theorem}
\usepackage[footnotesize]{caption}

\newcommand{\cleqn}{\setcounter{equation}{0}}
\newcommand{\newc}{\newcommand}
\newc{\be}{\begin{equation}}
\newc{\ee}{\end{equation}}
\newc{\bea}{\begin{eqnarray}}
\newc{\eea}{\end{eqnarray}}
\newc{\ol}{\overline}
\newc{\wt}{\widetilde}
\newc{\bs}{\boldsymbol}
\newc{\m}{\mathcal}

\begin{document}

\title{\hfill ~\\[-30mm]
       \hfill\mbox{\small SHEP-11-17}\\[30mm]
       \textbf{Trimaximal neutrino mixing from vacuum alignment in $\bs{A_4}$ and $\bs{S_4}$ models}}
\date{}
\author{\\Stephen F. King\footnote{E-mail: {\tt king@soton.ac.uk}}~~and
        Christoph Luhn\footnote{E-mail: {\tt christoph.luhn@soton.ac.uk}}\\ \\
  \emph{\small{}School of Physics and Astronomy, University of Southampton,}\\
  \emph{\small Southampton, SO17 1BJ, United Kingdom}}

\maketitle

\begin{abstract}
\noindent  
Recent T2K results indicate a sizeable reactor angle $\theta_{13}$
which would rule out {\em exact} tri-bimaximal lepton mixing. 
We study the vacuum alignment of the Altarelli-Feruglio $A_4$ family symmetry model including
additional flavons in the ${\bf 1'}$ and ${\bf 1''}$ representations and show that it
leads to trimaximal mixing in which the second column of the
lepton mixing matrix consists of the column vector  $(1,1,1)^T/\sqrt{3}$, with
a potentially large reactor angle.
In order to limit the reactor angle and control the higher order corrections, we propose a renormalisable
$S_4$ model in which the ${\bf 1'}$ and ${\bf 1''}$ flavons of $A_4$ are unified
into a doublet of $S_4$ which is spontaneously broken to $A_4$
by a flavon which enters the neutrino sector at higher order.
We study the vacuum alignment in the 
$S_4$ model and show that it predicts {\em accurate} trimaximal mixing with
{\em approximate} tri-bimaximal mixing, leading to a new mixing sum rule testable in future neutrino experiments.
Both $A_4$ and $S_4$ models preserve form dominance and hence predict zero leptogenesis, up to renormalisation group corrections.
 \end{abstract}
\thispagestyle{empty}
\vfill
\newpage
\setcounter{page}{1}

%%%%%%%%%%%%%%%%%%%%%%%%%%%%%%%%%%%%%%%%%

%%%%%%%%%%%%%%%%%%%%%%%%%%%%%%%%%%%%%%%%%

%%%%%%%%%%%%%%%%%%%%%%%%%%%%%%%%%%%%%%%%%
\section{Introduction}
\cleqn

Recently T2K have published evidence for a 
large non-zero reactor angle \cite{Abe:2011sj} which, when combined with data from MINOS and other
experiments in a global fit yields \cite{Fogli:2011qn},
\be
\theta_{13}=8^\circ\pm 1.5^\circ,
\ee
where the errors indicate the one $\sigma$ range, although the statistical significance of a non-zero
reactor angle is about 3$\sigma$.

If confirmed this would rule out the hypothesis
of exact tri-bimaximal (TB) mixing~\cite{Harrison:2002er}.
However other schemes such as trimaximal (TM) mixing remain viable 
\cite{Haba:2006dz,He:2006qd,Grimus:2008tt,Ishimori:2010fs,He:2011gb}:
\begin{equation}
\label{TM}
U_{\mathrm{TM}}~=~P'\begin{pmatrix} 
\frac{2}{\sqrt{6}}\cos\vartheta 
&\frac{1}{\sqrt{3}}
&\frac{2}{\sqrt{6}}\sin\vartheta \,e^{i\rho}\\
-\frac{1}{\sqrt{6}}\cos\vartheta -\frac{1}{\sqrt{2}}\sin\vartheta\,e^{-i\rho}
&\frac{1}{\sqrt{3}}
& \phantom{-}\frac{1}{\sqrt{2}}\cos\vartheta-\frac{1}{\sqrt{6}}\sin\vartheta  \,e^{i\rho} \\ 
-\frac{1}{\sqrt{6}}\cos\vartheta+ \frac{1}{\sqrt{2}}\sin\vartheta\,e^{-i\rho}
& \frac{1}{\sqrt{3}}
& - \frac{1}{\sqrt{2}}\cos\vartheta-\frac{1}{\sqrt{6}}\sin\vartheta \,e^{i\rho} 
\end{pmatrix} P \ ,
\end{equation}
where $\frac{2}{\sqrt{6}} \sin \vartheta=\sin\theta_{13}$, $P'$ is a diagonal phase matrix
required to put $U_{\mathrm{TM}}$ into the PDG convention \cite{Nakamura:2010zzi}, and
$P={\rm diag}(1,e^{i\frac{\alpha_{2}}{2} }, e^{i\frac{\alpha_{3}}{2} })$
contains the usual Majorana phases.
In particular TM mixing approximately preserves the successful TB mixing for the
solar mixing angle $\theta_{12}\approx 35^\circ$ as the correction due to a
non-zero but relatively small reactor angle is of second order.
Although TM mixing reduces to TB mixing in the limit that $\vartheta \rightarrow 0$,
it is worth emphasising that in general TM mixing involves a reactor angle
$\theta_{13}$ which could in principle be large or even maximal (e.g. $45^\circ$).
The observed smallness of the reactor angle~$\theta_{13}$ compared to
the atmospheric angle $\theta_{23}\approx 45^\circ$ and the solar angle
$\theta_{12}\approx 34^\circ$  \cite{Fogli:2011qn} 
is therefore not explained by the TM hypothesis alone. Clearly the relative smallness of the reactor angle
can only be explained with additional model dependent input. We shall show that, although an $A_4$ family
symmetry can provide
an explanation of trimaximal mixing, the observed smallness of the reactor angle as compared to the
atmospheric and solar angles may be naturally explained by an $S_4$ family symmetry. 

In the original $A_4$ models of TB mixing Higgs fields \cite{Ma:2001dn} or flavons
\cite{Altarelli:2005yp,Altarelli:2005yx} transforming under $A_4$
as ${\bf 3}$ and ${\bf 1}$ but not ${\bf 1'}$ or ${\bf 1''}$ were used to
break the family symmetry and to lead to TB mixing. 
However there is no good reason not to include flavons transforming as ${\bf
  1'}$ or ${\bf 1''}$ and once included
they will lead to deviations from TB mixing \cite{Brahmachari:2008fn,Ma:2011yi} in particular 
it has been noted that they lead to TM mixing \cite{Shimizu:2011xg}. However, as remarked above, TM 
mixing by itself does not account for the smallness of the reactor angle, and in particular the model in 
\cite{Shimizu:2011xg} does not provide any explanation for this. Another aspect is that higher order operators
in $A_4$ models lead to deviations from TB mixing in a rather uncontrolled way \cite{Altarelli:2005yx},
and in such models any simple TM structure may be washed out unless the theory is promoted to 
a renormalisable one in which the higher order operators are under control.
For example, this was done for two $A_4$ models of TB mixing in \cite{Varzielas:2010mp},
and one may follow an analogous strategy including also ${\bf 1'}$ or ${\bf 1''}$ flavons.

In this paper we make a first study of the vacuum alignment of an $A_4$ family
symmetry model of leptons with additional flavons in the  
${\bf 1'}$ and/or ${\bf 1''}$ representations and show that it leads to {\em
  accurate} trimaximal mixing at leading order (LO). 
In order to constrain the reactor angle and control the higher order corrections, 
we then propose a renormalisable 
$S_4$ model of leptons in which the ${\bf 1'}$ and ${\bf 1''}$ flavons of $A_4$ are unified
into a doublet of $S_4$ which is spontaneously broken to $A_4$
by a flavon which enters the neutrino sector at higher order.
We study the vacuum alignment of the $S_4$ model and show that 
it predicts {\em accurate} trimaximal mixing with {\em approximate} tri-bimaximal mixing, 
leading to a new mixing sum rule testable in future neutrino experiments.
We also remark that both $A_4$ and $S_4$ models preserve form dominance \cite{Chen:2009um}
and hence predict zero leptogenesis \cite{Choubey:2010vs}, up to renormalisation group corrections.

The layout of the rest of the paper is as follows.
In Section~\ref{sec-A4} we first revisit the Altarelli-Feruglio $A_4$ model \cite{Altarelli:2005yx}
with regard to the possibility of generating deviations from TB mixing which respect TM
mixing. As the obvious ideas do not yield TM mixing, we then consider the model augmented
by extra flavons in the ${\bf 1'}$ and/or ${\bf 1''}$ representation.
We study the vacuum alignment and show that this model leads to 
TM mixing, with an unconstrained reactor angle. 
In Section~\ref{sec-s4} we propose a renormalisable
$S_4$ model of leptons and study its vacuum alignment leading to 
{\em accurate} trimaximal mixing with {\em approximate} tri-bimaximal mixing.
In Section~\ref{sec-analytic} we give an analytic discussion of the
perturbative deviations to TB mixing arising from any TM model with a physical reactor angle 
leading to a new mixing sum rule testable in future neutrino experiments.
Section~\ref{sec-concl} concludes the paper. Appendix A contains 
Clebsch-Gordon coefficients for $S_4$ and $A_4$, and Appendix B describes
a perturbative diagonalisation of the right-handed neutrino mass matrix.

%%%%%%%%%%%%%%%%%%%%%%%%%%%%%%%%%%%%%%%%%

%%%%%%%%%%%%%%%%%%%%%%%%%%%%%%%%%%%%%%%%%

%%%%%%%%%%%%%%%%%%%%%%%%%%%%%%%%%%%%%%%%%

\section{\label{sec-A4}$\bs{A_4}$ models of trimaximal mixing}
\cleqn
\subsection{\label{sec-AF}The Altarelli-Feruglio ${\bs{A_4}}$ model of tri-bimaximal mixing}

The original and well studied Altarelli-Feruglio (AF) model of lepton masses
and mixings \cite{Altarelli:2005yx} is formulated as an effective theory,
defined purely by the particle content and the symmetries.  
There exist two versions, one with right-handed neutrinos and one without. For
definiteness we will only consider the former which makes use of the elegant
seesaw mechanism to obtain effective light neutrino masses around the eV scale.
The particle content and the symmetries of the AF model we consider 
is presented in Table~\ref{tab-AF}, including the $\xi$ flavon, but excluding the 
$\xi'$ and $\xi''$ flavons which we shall consider later.
In addition the vacuum alignment in the AF model requires a further auxiliary flavon singlet
$\tilde \xi$ which does not acquire a VEV and is not shown in the table.
Particles with a $U(1)_R$ charge of 2 are called driving fields. Setting their
$F$-terms to zero leads to the $F$-term conditions which control the alignment
of the flavon fields. 

The relevant effective superpotential terms of the Yukawa sector of the AF model are
\be
W_{A_4}^{\mathrm{eff}}~\sim~ LH_u N^c + (\varphi_{S} + \xi)N^cN^c  + \frac{1}{M} H_d\Big[ (L \varphi_T )_1 \,e^c +  (L \varphi_T
)_{1'} \,\mu^c +(L \varphi_T)_{1''} \,\tau^c \Big]\ .
\ee
Here $(\cdots)_{r}$ denotes the contraction of the $A_4$ indices to the
representation ${\bf{r}}$. When the three flavon fields $\varphi_{S},\xi,\varphi_T$
acquire their VEVs  \cite{Altarelli:2005yx},

\be
\langle \varphi_S \rangle = v_S 
\begin{pmatrix}1\\ 1\\1 \end{pmatrix} \ , \qquad\langle  \xi \rangle =u \ , \qquad 
\langle \varphi_T \rangle =v_T 
\begin{pmatrix}1\\ 0\\0 \end{pmatrix} \ ,\label{eq-AFalignment}
\ee
at some high scale, the flavour structure of the Yukawa couplings is generated
which, after electroweak symmetry breaking, gives rise to a trivial Dirac
neutrino mass matrix $m_D$, a TB heavy right-handed neutrino mass
matrix $M_R$ as well as a diagonal charged lepton mass matrix $m_\ell$.
Applying the seesaw formula yields a TB light neutrino mass matrix which is
diagonalised by $U_{\mathrm{TB}}=U_{\mathrm{TM}}(\vartheta=0)$. With the
charged leptons being already diagonal, the Pontecorvo-Maki-Nakagawa-Sakata
(PMNS) mixing matrix becomes $U_\mathrm{PMNS} = U_{\mathrm{TB}}$. 
The reason why the AF $A_4$ model preserves TB mixing is that the flavon $ \varphi_S$
preserves the $S$ generator of $A_4$ and the absence of $\xi'$ and $\xi''$ flavons leads to 
an accidental $U$ symmetry, where $S,U$ symmetry in the neutrino sector and $T$
symmetry in the charged lepton sector implies TB mixing (see e.g. \cite{King:2009ap}).

\begin{table}
$$
\begin{array}{cccccccccccccccc}\toprule
&N^c&L&e^c&\mu^c & \tau^c&H_{u}& H_{d}&\varphi_T&\varphi_S&\xi&\xi'& \xi''& \varphi_T^0&\varphi_S^0&\xi^0 \\  \midrule
A_4 & {\bf 3} & {\bf 3} & {\bf 1} & {\bf 1''} & {\bf 1'} & {\bf 1}  & {\bf 1} &{\bf 3} & {\bf 3} & {\bf 1} & {\bf 1'} & {\bf 1''} & {\bf 3} & {\bf 3} & {\bf 1}  \\[2mm]
Z_3 & \omega^2 & \omega & \omega^2 & \omega^2 & \omega^2 &1&1&1& \omega^2 &\omega^2 & \omega^2 & \omega^2 &1& \omega^2 & \omega^2 \\[2mm]
U(1)_R&1&1&1&1&1&0&0&0&0&0&0&0&2&2&2 \\ \bottomrule
\end{array}
$$
\caption{\label{tab-AF}The particle content and symmetries of the
  $A_4$ model with extra $\xi'$, $\xi''$ flavons.}
\end{table}

This result is correct at LO. It is argued in
\cite{Altarelli:2005yx} that higher order operators  which are allowed by the
imposed symmetries should be considered as well, both in $W_{A_4}^{\mathrm{eff}}$ as well as in
the superpotential that generates the flavon alignments in
Eq.~\eqref{eq-AFalignment}. In general such additional terms would lead to
deviations from  TB mixing which are important in the light of
the latest experimental hints for a non-zero reactor angle \cite{Abe:2011sj,Fogli:2011qn}. 
In order to account for a (relatively) large value of $\theta_{13}$ the higher
order corrections need to be large enough which can be arranged by assuming
different suppression scales for different non-renormalisable operators.
This assumption, however, raises the question about the origin of the
suppression scales.

In this context, it was pointed out recently that particular ultraviolet (UV)
completions of flavour models do not necessarily give rise to all effectively
allowed terms in the superpotential \cite{Varzielas:2010mp}. In fact, it
was found that the minimal UV completion of the AF model does not generate any
deviations from TB mixing. The simplest way to obtain the desired
deviations is to add new messengers to the minimal model such that certain
higher order Yukawa operators are switched on. In the neutrino sector, the
effective non-trivial operators allowed by the symmetries read\footnote{Note
  that the third term, corresponding to two independent contractions, was not
  listed in Eq.~(32) of \cite{Altarelli:2005yx} despite giving non-trivial
  contributions to the mass matrix, i.e. contributions that differ from the
  tri-bimaximal structure.}  
\be
\frac{1}{M}\Big[(N^cN^c)_{1'} (\varphi_S\varphi_T)_{1''} + 
(N^cN^c)_{1''} (\varphi_S\varphi_T)_{1'} + 
(N^cN^c)_{3} (\varphi_S\varphi_T)_{3_{s,a}} + 
(N^cN^c)_{3} (\xi \varphi_T)_{3}  \Big] ,\label{AF-completion}
\ee
all of which break TB neutrino mixing. Inspection of the flavour
structure of these terms shows that only the first two terms lead to mass
matrices that have an eigenvector $\frac{1}{\sqrt{3}}(1,1,1)^T$, thus retaining
TM neutrino mixing. However, requiring the messengers that mediate these
effective operators to be matter-like, i.e. they should have a $U(1)_R$ charge
of 1, we find that we will always get a contribution from the third term of
Eq.~\eqref{AF-completion} as well. Therefore, without introducing new flavon
fields, we cannot find a simple UV completion of the AF model
where the neutrino mixing features a trimaximal pattern while breaking the
tri-bimaximal one.

%%%%%%%%%%%%%%%%%%%%%%%%%%%%%%%%%%%%%%%%%%
%%%%%%%%%%%%%%%%%%%%%%%%%%%%%%%%%%%%%%%%%%

\subsection{\label{sec-xiprime-struc}Trimaximal mixing from $\bs{A_4}$ with extra $\bs{\xi'}$ and  $\bs{\xi''}$ flavons}

The discussion in the previous subsection leads us to consider the case
where extra flavons $\xi'$ in the ${\bf 1'}$ representation and $\xi''$ in the
${\bf 1''}$ representations  of $A_4$ are added to the model as already shown
in Table~\ref{tab-AF}.  This has previously been suggested (without the
see-saw mechanism) in \cite{Shimizu:2011xg} where the phenomenological
consequences of the LO terms were studied numerically.
However the flavon alignment was not derived in \cite{Shimizu:2011xg} but simply postulated.
Remarkably, although the difference between the $\xi'$ and $\xi''$ flavon VEVs
breaks the  accidental $U$ symmetry and thereby violates TB mixing, the presence
of these flavons respects the $S$ symmetry and leads to TM mixing.

In this subsection we consider the effect on the neutrino mass matrices of adding
flavons $\xi'$ in the ${\bf 1'}$ representation and/or $\xi''$ in the ${\bf
  1''}$ representations. In the subsequent subsection we consider the
vacuum alignment problem including these flavons. 
Assuming the flavon alignments in Eq.~\eqref{eq-AFalignment}, it is straightforward to find the structure of
the charged lepton and the light neutrino mass matrices. As the charged lepton
Yukawa couplings are non-renormalisable, a particular set of messengers is
necessary to generate the required couplings. Following
\cite{Varzielas:2010mp}, one can show that a minimal messenger completion does
not generate any off-diagonal entries in $m_\ell$. Therefore, the leptonic
mixing matrix $U_{\mathrm{PMNS}}$ is solely determined by the neutrino
sector. Given the  symmetries of Table~\ref{tab-AF}, the corresponding
renormalisable neutrino part of the superpotential is extended to,
\be
W^{\nu}_{A_4} ~=~ y LH_u N^c + (y_1 \varphi_S + y_2 \xi +y'_3 \xi' +y''_3 \xi'')N^cN^c   \ .\label{Ynu}
\ee
Inserting the flavon vacuum alignments in Eq.~\eqref{eq-AFalignment}, and
assuming both $\xi'$ and $\xi''$ as well as SM Higgs VEVs, we obtain the Dirac
neutrino mass matrix $m_D$ as well as the right-handed neutrino mass matrix $M_R$,
\be
m_D~=~\begin{pmatrix}1&0&0\\0&0&1\\0&1&0 \end{pmatrix} y \,v_u \ ,\label{mD}
\ee
\be
M_R~=~\left[  \alpha \begin{pmatrix} 2&-1&-1\\-1&2&-1\\-1&-1&2\end{pmatrix}
+\beta  \begin{pmatrix}1&0&0\\0&0&1\\0&1&0 \end{pmatrix}
+\gamma' \begin{pmatrix}0&0&1\\0&1&0\\1&0&0 \end{pmatrix}
+\gamma''  \begin{pmatrix} 0&1&0\\1&0&0\\0&0&1\end{pmatrix}  \right] 
\ ,\label{mR}
\ee
where $\alpha= y_1v_S$, $\beta= y_2  \langle  \xi \rangle$, 
$\gamma'=y_3' \langle  \xi' \rangle$,  $\gamma''=y_3'' \langle  \xi'' \rangle$.

The complex symmetric matrix $M_R$ is
diagonalised by a unitary transformation $U_R$,
\be
U_R^T M^{}_R U_R^{} ~=~ M_R^{\mathrm{diag} } \ .
\ee
One can easily verify that $\frac{1}{\sqrt{3}}(1,1,1)^T$ is an eigenvector of
$M_R^\dagger M_R^{}$, the eigenvalue being $|(\beta+\gamma'+\gamma'')|^2$. Therefore the
matrix $U_R$ will have a trimaximally mixed column which, for phenomenological
reasons, will be identified as the second column. The other two columns are
more difficult to determine and we will discuss the analytic approximate
calculation in Section \ref{sec-analytic}. However, they can be parameterised
as a linear combination of the first and third columns of the TB
mixing matrix $U_{\mathrm{TB}}$. Thus $U_R$ is of the form given in Eq.~\eqref{TM},
with the parameters $\vartheta$ and $\rho$ being functions of the
Yukawa couplings and VEVs. Note that a Majorana phase matrix would still have to be multiplied on
the right in order to render the right-handed neutrino masses
real. Disregarding this and other issues about phase conventions, we can now
rewrite the effective light neutrino mass matrix,
\be
m_\nu^{\mathrm{eff}}
= 
- m_D^{} M_R^{-1} m_D^T
=
-m_D^{} U_R^{}  (M^{\mathrm{diag}}_R)^{-1} U_R^T m_D^T
=
- U_\nu^{}\,  \frac{(y\,v_u)^2}{M^{\mathrm{diag}}_R} \,U_\nu^T  \ ,\label{meff}
\ee
where the matrix $U_\nu$ is defined as
\be
U_\nu~=~\frac{m_D}{y v_u} \, U_R  \ .\label{URUnu}
\ee
Due to the trivial structure of $m_D$, see Eq.~\eqref{mD}, it is unitary
and again of trimaximal form. In fact, $U_\nu$ differs from $U_R$ only in the
interchange of the second and the third row. 
Assuming diagonal charged leptons, the matrix $U_\nu$ is just the lepton mixing
matrix $U_\nu = U_{\mathrm{PMNS}}$. From Eq.~\eqref{meff} it is then apparent
that the Dirac neutrino mass matrix in the basis of diagonal right-handed neutrinos and charged leptons,
is given by $m'_D=yv_uU_{\mathrm{PMNS}}$. In this basis, the columns of $m'_D$ are proportional
to the columns of $U_{\mathrm{PMNS}}$, and thus we see that 
the $A_4$ model with additional $\xi'$ and $\xi''$ flavons satisfies form dominance
\cite{Chen:2009um} and hence leptogenesis vanishes
\cite{Choubey:2010vs}.\footnote{This conclusion can be
  avoided by taking into account renormalisation group corrections which effect the
  entries of the Dirac neutrino Yukawa matrix differently so that the
  orthogonality of the columns is broken \cite{cooper}.}
However, form dominance does not imply that  
the effective light neutrino mass matrix is
form diagonalisable, and indeed this is not the case since
the mixing matrix $U_\nu$ depends on the 
Yukawa couplings and VEVs (which in turn determine the neutrino masses). Only in the
limit where $\gamma'=\gamma''$  do we recover a form diagonalisable mass matrix
corresponding to the TB case. 

The size of the deviations from TB
mixing are controlled by the parameters $y_3'$ and  $y_3''$ as well as the
VEVs $ \langle \xi' \rangle$ and $ \langle \xi'' \rangle$. From the vacuum
alignment discussion in the next subsections,  however, there is no reason to
believe that these VEVs should be small, i.e. one would expect a large reactor angle
as well as large deviations from maximal atmospheric mixing. In order to meet
the experimental bounds, one would therefore have to assume the input
parameters of the flavon and Yukawa superpotential to conspire so as to yield
only small deviations from the TB mixing pattern. Eventually this will motivate us to 
go beyond $A_4$ to $S_4$ where the two extra flavons $\xi'$ and $\xi''$ are unified
into a doublet representation.

%%%%%%%%%%%%%%%%%%%%%%%%%%%%%%%%%%%%%%%%%%

\subsection{\label{sec-xiprime-align}Vacuum alignment in $\bs{A_4}$ with extra
  $\bs{\xi'}$ and $\bs{\xi''}$ flavons}

In this subsection we scrutinise the modifications to the alignment mechanism of
Altarelli and Feruglio due to the addition of flavons $\xi'$ in the ${\bf 1'}$
and/or  $\xi''$ in the ${\bf 1''}$ representation of $A_4$. 
Our main result is that contrary to
\cite{Altarelli:2005yx} one need not introduce the auxiliary flavon singlet
$\tilde \xi$ which does not acquire a VEV and so plays no direct role in the
flavour structure of the neutrinos. In that sense the introduction of a $\xi'$
and/or $\xi''$
flavon does not only serve the purpose of breaking the TB structure to a TM
one, but also does not complicate the model at all.

The renormalisable terms of the driving superpotential, replacing the $\tilde
\xi$ flavon of the AF model by the $\xi'$ and $\xi''$ flavons, are,
\bea
W^{\mathrm{flavon}}_{A_4} &=& \varphi_T^0 \left(M\varphi_T + g \varphi_T \varphi_T
\right)  + \varphi_S^0 \left(g_1 \varphi_S \varphi_S + g_2 \varphi_S \xi +
g'_3\varphi_S \xi' + g''_3\varphi_S \xi'' \right)  \notag \\[2mm]
&&
 +  \xi^0 \left( g_4 \varphi_S\varphi_S + g_5 \xi \xi +g_6 \xi' \xi'' \right)    .\label{a4-flavon}
\eea
As the $F$-term equations of $\varphi_T^0$ remain unchanged from the AF model, we
obtain - up to some $A_4$ transformed solutions -  the well known alignment
\be
\langle \varphi_T \rangle = v_T \begin{pmatrix}1\\0\\0 \end{pmatrix} , \qquad
v_T = -\frac{M}{2g} \ .\label{phiT-alignment}
\ee
On the other hand, the $F$-term conditions of $\varphi_S^0$ and $\xi^0$ are
slightly modified in this new setup. Writing $\langle {\varphi_S}_i \rangle =
s_i$ and  $\langle {\xi} \rangle =u$, $\langle {\xi'} \rangle =u'$ and
$\langle {\xi''} \rangle =u''$,we get 
\be
2g_1\begin{pmatrix} s_1^2-s_2s_3\\s_3^2-s_1s_2\\s_2^2-s_3s_1 \end{pmatrix} 
+
g_2 u \begin{pmatrix} s_1\\s_2\\s_3 \end{pmatrix} 
+
g'_3 u' \begin{pmatrix} s_3\\s_1\\s_2 \end{pmatrix} 
+
g''_3 u'' \begin{pmatrix} s_2\\s_3\\s_1 \end{pmatrix} 
= \begin{pmatrix} 0\\0\\0 \end{pmatrix} ,\label{ftermAF}
\ee
\be
g_4 (s_1^2+2s_2s_3) +g_5 u^2 + g_6 u' u''~=~ 0 \ .\label{ftermAF-2}
\ee

For simplicity, let us first consider the case where only one of the two non-trivial
one-dimensional flavons is present. For definiteness we assume 
this to be the $\xi'$.
It is then straightforward to work out the most general solutions to 
Eqs.~(\ref{ftermAF},\ref{ftermAF-2}). Again, disregarding the ambiguity caused
by $A_4$ symmetry transformations we find two possible non-trivial solutions,
\be
\langle \varphi_S \rangle = v_S \begin{pmatrix}1\\1\\1 \end{pmatrix} ,
\qquad v_S^2 =  -\frac{g_5}{3g_4} u^2   \ , 
\qquad u' = -\frac{g_2}{g_3'} u \ ,\label{phiS-alignment}
\ee
as well as
\be
\langle \varphi_S \rangle = v_S \begin{pmatrix}0\\1\\0 \end{pmatrix} ,
\qquad  v_S = - \frac{g'_3} {2g_1}u' \ ,
\qquad u = 0 \ .
\ee
Choosing the soft mass parameter $m^2_{\xi}<0$, the second solution is
eliminated and the VEV $u$ slides to a large scale \cite{Altarelli:2005yx}. In
this way we are able to get the original alignment of $\varphi_S$ and
non-vanishing VEVs for $\xi$ and $\xi'$, see Eq.~\eqref{phiS-alignment}.

We now consider the effect of having both flavons  $\xi'$ and $\xi''$ in the
${\bf 1'}$ and ${\bf 1''}$  representations of $A_4$. Then the terms proportional
to $g_3''$ and $g_6$ would be switched on in the flavon superpotential of
Eq.~\eqref{a4-flavon} The corresponding extra terms in the $F$-term conditions
of Eqs.~(\ref{ftermAF},\ref{ftermAF-2}) would thus modify the physical
solution of Eq.~\eqref{phiS-alignment} to 
\be
\langle \varphi_S \rangle = v_S \begin{pmatrix}1\\1\\1 \end{pmatrix} ,
\qquad v_S^2 =  -\frac{g_5u^2 +g_6 u'u''}{3g_4}   \ , 
\qquad u = -\frac{g'_3 u'+g''_3 u''}{g_2} \ .\label{phiS-alignment-mod}
\ee
This solution has the unpleasant feature of leading to arbitrary physics. For
instance, if $y'_3  u' =  y''_3 u''$, where $y''_3$ denotes the Yukawa
coupling of $\xi''$ to $N^cN^c$, see Eq.~\eqref{Ynu}, then this implies that 
the mass matrix $M_R$ in Eq.~\eqref{mR} has a tri-bimaximal structure, since $\gamma'=\gamma''$, and thus
the reactor angle vanishes identically.
It is for this reason that models with {\it either} $\xi'$
{\it or} $\xi''$ essentially yield the same physics. However, adding
both types of flavons in $A_4$ generates a bothersome ambiguity in physical predictions.
In the next section the above ambiguity is removed by unifying the flavons $\xi'$ and $\xi''$ 
into an $S_4$ doublet $\eta_{\nu}$, whose VEV components are aligned along a $U$ preserving direction,
thereby restoring TB mixing, at least approximately.

%%%%%%%%%%%%%%%%%%%%%%%%%%%%%%%%%%%%%%%%%

%%%%%%%%%%%%%%%%%%%%%%%%%%%%%%%%%%%%%%%%%

%%%%%%%%%%%%%%%%%%%%%%%%%%%%%%%%%%%%%%%%%

\section{\label{sec-s4}${\bs{S_4}}$ model of trimaximal mixing}
\cleqn

\subsection{The effective $\bs{S_4}$ model of leptons}
As pointed out in the previous section, the $A_4$ model with $\xi'$ and/or
$\xi''$ flavons cannot explain the smallness of the deviations from TB
mixing. Furthermore, adding both non-trivial one-dimensional flavons leads to
an ambiguity in physical predictions. In order to cure these shortcomings 
we consider an $S_4$ model in which the ${\bf 1'}$ and ${\bf 1''}$ 
representations of $A_4$ are unified into the $\eta_{\nu}$ doublet of
$S_4$ while the triplet representations remain.
The complete list of lepton, Higgs and flavon fields is given in
Table~\ref{tab-S4}.\footnote{ In principle the triplet fields can either be identified
with the ${\bf 3}$ or the ${\bf 3'}$ of $S_4$. They differ from each other
only in the sign of the $U$ generator (see Appendix~\ref{a4-s4-CGs}) such that
all representation matrices of the ${\bf 3}$ have determinant $+1$, while this
is not the case for the~${\bf 3'}$. In the case of the right-handed neutrinos
$N^c$ and the lepton doublet $L$, we are free to choose the type of $S_4$
triplet as long as it is the same for both fields, and we choose the ${\bf 3}$.
Note that the triplet flavon $\varphi_\nu$ must furnish a ${\bf 3'}$  of $S_4$
  because it is coupled to the {\it symmetric} product $N^cN^c$.} 
Similar to the $A_4$ model we have a $U(1)_R$ symmetry as well as a $Z_3$
symmetry which separates the neutrino and the charged lepton sector. 

\begin{table}
$$
\begin{array}{ccccccccccccccc}\toprule
&N^c&L&e^c&\mu^c & \tau^c
&H_{u}&H_{d}
&\varphi_\ell&\eta_\mu&\eta_e
&\varphi_\nu&\eta_\nu&\xi_\nu&\zeta_\nu\\  
\midrule
S_4 & {\bf 3} & {\bf 3} & {\bf 1} & {\bf 1} & {\bf 1'} 
& {\bf 1}  & {\bf 1}
&{\bf 3'} & {\bf 2} & {\bf 2} 
& {\bf 3'} & {\bf 2} & {\bf 1} & {\bf 1'}
\\[2mm]
Z_3 & \omega^2 & \omega & \omega^2 & \omega^2 & \omega^2 
&1&1
&1& 1 &1 
& \omega^2 &\omega^2& \omega^2 & 1 \\[2mm]
Z'_3 & 1 & 1 &1 & \omega & \omega^2 
&1&1
& \omega&  \omega & \omega^2 
& 1 &1&1 & 1 \\[2mm]
U(1)_R&1&1&1&1&1&0&0&0&0&0&0&0&0&0 \\ \bottomrule
\end{array}
$$
\caption{\label{tab-S4}Lepton, Higgs and flavon fields of the $S_4$ model.}
\end{table}

In the neutrino sector of the $S_4$ model there are three flavon fields:
 $\varphi_\nu$ and $\xi_\nu$
(analogous to $\varphi_S$ and $\xi$
  of the AF model) and $\eta_\nu$ (which unifies the two $A_4$ flavon fields $\xi'$ and $\xi''$).
The neutrino part of the effective superpotential is then,
\bea
W_{S_4}^{\nu, \mathrm{eff}} &\sim & LH_u N^c + ( \varphi_\nu  + \xi_\nu+  \eta_\nu)N^cN^c
+ \frac{\zeta_\nu }{M_\chi}\,\eta_\nu N^cN^c ,
\label{s4-Ynu0}
\eea
analogous to Eq.~\eqref{Ynu}, where, as in the $A_4$ model, the Dirac neutrino
mass matrix takes the trivial form $m_D$ of Eq.~\eqref{mD}. However 
an additional flavon $\zeta_\nu$ in the ${\bf 1'}$
representation of $S_4$ serves the purpose of breaking $S_4$ to $A_4$. 
The $U$ violating flavon $\zeta_{\nu}$ is forbidden by 
$S_4$ to couple to $N^cN^c$ at renormalisable level but 
appears at higher order, thereby breaking the TB structure by a small amount,
leading to a non-zero but suppressed reactor angle,
while preserving the TM structure due to the unbroken $S$ symmetry. 
Inserting the flavon VEVs, whose alignment is discussed later,
\be
\langle \varphi_\nu \rangle = v_\nu \begin{pmatrix} 1\\1\\1 \end{pmatrix} , 
\qquad
\langle\eta_\nu  \rangle = w_\nu \begin{pmatrix} 1\\1 \end{pmatrix} , 
\qquad
\langle \xi_\nu \rangle = u_\nu \ ,
\qquad
\langle \zeta_\nu \rangle = z_\nu \ ,\label{s4-align-nu}
\ee
$W_{S_4}^{\nu, \mathrm{eff}}$ leads to
TB neutrino mixing at LO, broken to TM mixing by the non-vanishing
VEV for the $\zeta_\nu$ flavon at next-to-leading order (NLO).
Note that the neutrino sector above may readily be incorporated without change
into an 
$SU(5)$ GUT as was done for $A_4$ and $S_4$ models of TB mixing along the lines of
\cite{Burrows:2009pi,Burrows:2010wz,Hagedorn:2010th,Antusch:2011sx}.

In order to account for the charged lepton mass hierarchy
we identify the right-handed charged leptons with
one-dimensional representations of $S_4$, and distinguish them
by an extra $Z'_3$ family symmetry, broken by a triplet flavon $\varphi_\ell$
as well as the doublet flavons $\eta_\mu,\eta_e$, resulting in the effective charged
lepton superpotential, 
\be
W_{S_4}^{\ell, \mathrm{eff}} ~\sim ~ \frac{\varphi_\ell}{M_{\Omega_1}} H_d \left[ L  \tau^c
+ \frac{ \eta_\mu}{M_{\Omega_2}}L \mu^c 
+ \frac{ \eta_e}{M_{\Omega_3}}L   e^c
\right]
 \ ,\label{s4-yuk-eff}
\ee
where $M_{\Omega_i}$ are {\it a priori} independent mass scales, although two of the mass scales will turn out to be
equal, $M_{\Omega_2} = M_{\Omega_3}$.
Inserting the flavon VEVs, 
\be
\langle \varphi_\ell \rangle = v_\ell 
\begin{pmatrix} 0\\1\\0 \end{pmatrix} \ , \qquad
\langle \eta_\mu \rangle =w_\mu 
\begin{pmatrix} 0\\1 \end{pmatrix} \ ,\qquad
\langle \eta_e \rangle =w_e 
\begin{pmatrix} 1\\0 \end{pmatrix}
 \ ,\label{s4-align-char0}
\ee
whose alignment is discussed later, and using the Clebsch-Gordan
coefficients listed in Appendix~\ref{a4-s4-CGs}, leads to
\be
W_{S_4}^{\ell, \mathrm{eff}} ~\sim ~ \frac{v_\ell H_d}{M_{\Omega_1}} \left[ L_3 \tau^c 
+\frac{ w_\mu }{M_{\Omega_2}} L_2 \mu^c 
+ \frac{w_e}{M_{\Omega_3}}L_1 e^c 
\right] \ , \label{hierarchy}
\ee
resulting in a diagonal charged lepton mass matrix $m_\ell$.
The larger $\tau$ lepton mass is explained
by the fact that the $e$ and $\mu$ masses are only provided at higher order
by two additional doublet flavons $\eta_\mu,\eta_e$.
The electron to muon mass hierarchy is accounted for by a hierarchy of flavon VEVs
$w_e \ll w_{\mu}$,  
which can be obtained consistently assuming certain
hierarchies in the mass parameters of the flavon superpotential. Such large
hierarchies of VEVs are of course stable in SUSY 
models, and are already familiar due to the hierarchy between the weak scale and the 
flavour or GUT scales. Finally note that 
the charged lepton sector above will eventually require modification 
as was done for $A_4$ and $S_4$ models of TB mixing
in order to be incorporated into an  
$SU(5)$ GUT since quark mixing requires some small off-diagonal mixing also in the charged lepton sector
\cite{Burrows:2009pi,Burrows:2010wz,Hagedorn:2010th,Antusch:2011sx}.

\subsection{The renormalisable $\bs{S_4}$ model of leptons }

As emphasised earlier, any non-renormalisable term of an effective
superpotential should be understood in terms of a more fundamental underlying
renormalisable theory. Without such a UV completion of a model, higher order
terms which are allowed by the symmetries may or may not be present. Thus a
purely effective formulation would  leave room for different physical
predictions. In order to remove any such ambiguity within our $S_4$ model
we have constructed a fully renormalisable theory of the lepton sector. The
required messengers  are listed in Table~\ref{tab-S4-2}, together with the
driving fields which control the alignment of the flavons.

\begin{table}
$$
\begin{array}{cccccccccccccc}\toprule
&\varphi_\ell^0&\eta_\ell^0&\xi_\ell^0
&\varphi_\nu^0&\tilde\varphi_\nu^0&\xi_\nu^0  &D^0
&\chi&\chi^c&\Omega_1^{}&\Omega_1^c&\Omega_2^{}&\Omega_2^c
\\  \midrule
S_4 & {\bf 3'} & {\bf 2} & {\bf 1} 
& {\bf 3'} & {\bf 3} & {\bf 1}  & {\bf 1} 
&{\bf 3'} & {\bf 3'} & {\bf 3} & {\bf 3} & {\bf 2} & {\bf 2}   \\[2mm]
Z_3 & 1&1&1
&\omega^2 & \omega^2 & \omega^2 & 1 
& \omega & \omega^2& \omega^2 &\omega^1 & \omega^2 & \omega  \\[2mm]
Z'_3 & \omega&\omega&\omega
&1&1&1&1
&1&1&1&1&\omega^2 & \omega  \\[2mm]
U(1)_R&2&2&2&2&2&2&2&1&1&1&1&1&1 \\ \bottomrule
\end{array}
$$
\caption{\label{tab-S4-2}Driving fields and messengers of the $S_4$ model.}
\end{table}

\begin{figure}
\begin{center}
\includegraphics[height=3cm]{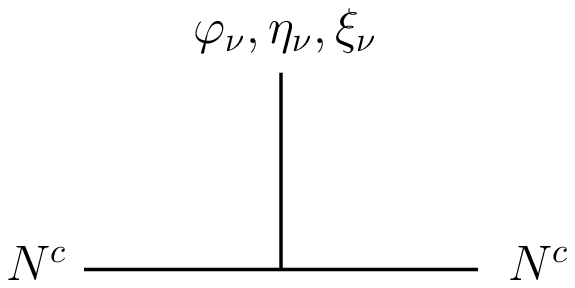}
~~~~~~
\includegraphics[height=3cm]{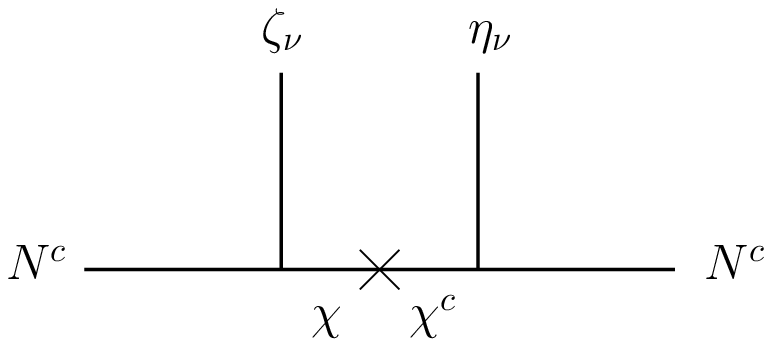}
\end{center}
\caption{\label{fig-seesaw}Leading and next-to-leading order right-handed
  neutrino mass contributions.}
\end{figure}

With the particle content and the symmetries specified in Tables~\ref{tab-S4}
and~\ref{tab-S4-2}, we can replace the effective neutrino superpotential in Eq.~\eqref{s4-Ynu0}
by the sum of two renormalisable pieces, a LO piece and a messenger piece,
\be
W^{\nu}_{S_4}=W^{\nu, \mathrm{ LO}}_{S_4} + W^{\nu, \mathrm{mess}}_{S_4} \ ,
\ee
where,
\bea
W^{\nu,\mathrm{LO}}_{S_4} &=& y LH_u N^c + (y_1 \varphi_\nu  +y_2 \xi_\nu+ y_3
\eta_\nu)N^cN^c \ ,
\label{s4-Ynu}\\[2mm]
W^{\nu, \mathrm{mess}}_{S_4} &=&  x_1 N^c \zeta_\nu \chi + \chi^c (x_2 \eta_\nu+x_3\varphi_\nu) N^c + M_\chi \chi
\chi^c \ . \label{s4-numess} 
\eea
The leading order contribution to the right-handed neutrino mass matrix $M_R$
is given by $W^{\nu,\mathrm{LO}}_{S_4}$ and leads to a TB structure.  Given a 
non-vanishing VEV for the $\zeta_\nu$ flavon, $W^{\nu, \mathrm{mess}}_{S_4} $
is responsible for breaking the TB structure to a TM one at higher order. 
The corresponding LO and NLO diagrams are depicted in
Figure~\ref{fig-seesaw}.\footnote{We remark that we have suppressed the additional 
 messenger terms $(\varphi_\nu + \eta_\nu + \xi_\nu)\chi^c\chi^c$ 
which are allowed in $W^{\nu, \mathrm{mess}}_{S_4}$
since they become relevant only at next-to-next-to-leading order with the
$\zeta_\nu$ flavon entering quadratically; as the Klein symmetry $Z_S\times
Z_U$ of the neutrino sector \cite{King:2009ap} is restored  in this diagram by
the quadratic appearance of $\zeta_\nu$, such a higher order term yields a TB
contribution to $M_R$. Therefore, the only significant term contributing to
the breaking of TB to TM is the one shown in Figure~\ref{fig-seesaw}.}
Integrating out the messenger pair $\chi,\chi^c$, the NLO diagram reproduces
uniquely the effective operator in Eq.~\eqref{s4-Ynu0}, 
$\frac{\zeta_\nu }{M_\chi}\,\eta_\nu N^cN^c$. Note that a similar term with
  $\eta_\nu$ replaced by $\varphi_\nu$ is forbidden by $S_4$ as the symmetric
  product  $N^cN^c$ does not include the triplet ${\bf 3}$.
Plugging in the flavon VEVs
and using the $S_4$ Clebsch-Gordan coefficients of Appendix~\ref{a4-s4-CGs}
we obtain a contribution to $M_R$ of the form
\be
\Delta M_R ~=~ 
x_1x_2\,\frac{w_\nu z_\nu }{M_\chi} \left[- \begin{pmatrix} 0&0&1\\0&1&0\\1&0&0  \end{pmatrix} +
\begin{pmatrix} 0&1&0\\1&0&0\\0&0&1  \end{pmatrix}
\right].
\label{DeltaMR}
\ee
The relative minus sign arises as the $\zeta_\nu$  VEV  breaks the $U$
symmetry of $S_4$. Its presence leads to a deviation from the TB structure
which would exist if the two matrices were added instead. 

The $S_4$ model thus gives rise to TB neutrino mixing at LO which is
broken to a TM mixing pattern by NLO corrections induced by
the VEV of the $S_4$ breaking flavon field~$\zeta_\nu$. This naturally
explains why the reactor angle as well as the deviations from maximal
atmospheric mixing are relatively small. Provided the charged leptons are
diagonal, the lepton mixing matrix is purely determined by the structure of
the right-handed neutrino mass matrix $M_R$, given in Eq.~\eqref{mR}, with
\be
\alpha=y_1 v_\nu \ , \qquad
\beta=y_2 u_\nu \ , \qquad
\gamma'= \gamma - \Delta \ , \qquad
\gamma''= \gamma + \Delta\ ,
\ee
where we have defined  
\be
\gamma=y_3 w_\nu \ , \qquad \Delta = x_1x_2 \frac{w_\nu z_\nu}{M_\chi} \ .
\label{inputs}
\ee
Notice that $\gamma'$ and $\gamma''$ are equal at LO. The deviations from a TB mass
matrix arise only at NLO which is parameterised by $\frac{\Delta}{\gamma} \sim \frac{z_\nu}{M_\chi}$.

%%%%%%%%%%%%%%%%%%%%%%%%%%%%%%%%%%%5

%%%%%%%%%%%%%%%%%%%%%%%%%%%%%%%%%%%5

%%%%%%%%%%%%%%%%%%%%%%%%%%%%%%%%%%%5

\begin{figure}
\begin{center}
\includegraphics[height=2.8cm]{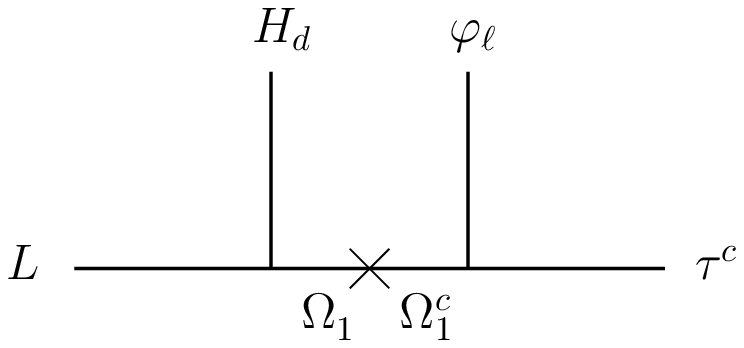}
~~~~~~
\includegraphics[height=2.8cm]{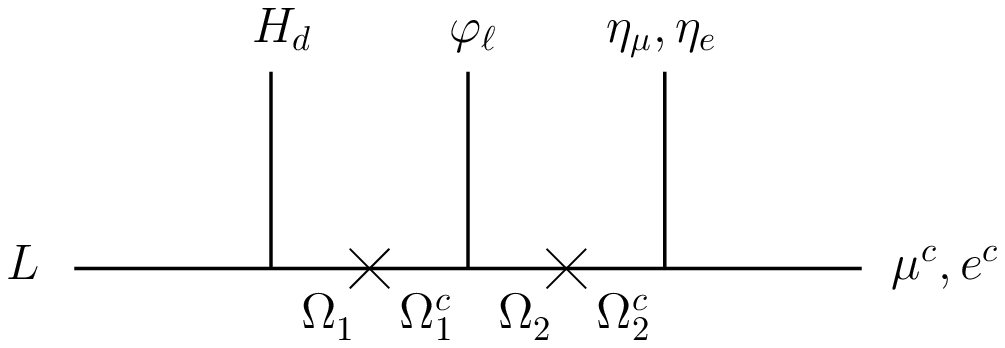}
\end{center}
\caption{\label{fig-charged}$S_4$ diagrams that lead to a diagonal charged
  lepton sector.} 
\end{figure}

The charged lepton sector is formulated at the renormalisable level,
using two new pairs 
of messengers, $\Omega_i^{}$ and $\Omega_i^c$ ($i=1,2$).  
With the particles and symmetries listed in Tables~\ref{tab-S4} and
\ref{tab-S4-2}, we get the renormalisable superpotential for the charged leptons,
\bea
W_{S_4}^{\ell} &\sim& 
LH_d \Omega_1^{} + \Omega_1^c \varphi_\ell \tau^c 
+  \Omega_1^c \varphi_\ell \Omega_2^{} 
+ \Omega_2^c \eta_\mu \mu^c
 + \Omega_2^c \eta_e e^c \notag \\[2mm]
&& + M_{\Omega_1} \Omega^{}_1 \Omega_1^c 
+ (M_{\Omega_2} + \zeta_\nu )\Omega^{}_2 \Omega_2^c  \ , \label{s4-yuk}
\eea
where we have suppressed all order one coupling constants. The appearance of
$\zeta_\nu$ in the second line leads to an irrelevant correction of the
$\Omega_2^{}$ messenger mass which we ignore in the following. Integrating out
the messengers, we obtain the diagrams of Figure~\ref{fig-charged}
leading uniquely to $W_{S_4}^{\ell, \mathrm{eff}}$ in Eq.~\eqref{s4-yuk-eff},
with $M_{\Omega_2}=M_{\Omega_3}$.

\subsection{\label{sec-s4-vac}Vacuum alignment in the $\bs{S_4}$ model}

So far we have only postulated the particular alignments of the neutrino-type
flavons, given in Eq.~\eqref{s4-align-nu}, and the flavons of the charged
lepton sector, Eq.~\eqref{s4-align-char0}. In this subsection we first explore the
driving potential relevant for the neutrino sector and prove that the assumed
alignment can in fact be obtained in an elegant way which is similar to the
$A_4$ case. Then we perform a similar study of the flavon alignments for  the
charged leptons.

The renormalisable superpotential involving the driving fields
necessary for aligning the neutrino-type flavons is given as
\bea
W_{S_4}^{\mathrm{flavon},\nu}&=& 
\varphi_\nu^0 (g_1 \varphi_\nu\varphi_\nu  
+ g_2 \varphi_\nu \xi_\nu+g_3 \varphi_\nu \eta_\nu)
+ \xi^0 (g_4 \varphi_\nu\varphi_\nu +  g_5 \xi_\nu\xi_\nu
+\frac{g_6}{2} \eta_\nu \eta_\nu ) \notag \\[2mm]
&&
+ g_7 \tilde \varphi_\nu^0   \varphi_\nu \eta_\nu  
+ D^0 (f_1 \zeta_\nu\zeta_\nu + f_2 \eta_\mu \eta_e + f_3 H_uH_d +f_4
M_0^2) \ .
\label{s4-flavon-nu}
\eea
Identifying the two components of the doublet flavon $\eta_\nu$ with $\xi''$ and
$\xi'$, the first line of Eq.~\eqref{s4-flavon-nu} is identical to the
corresponding terms in the $A_4$ driving superpotential of
Eq.~\eqref{a4-flavon}. Due to the $S_4$ structure, we additionally get
$g_3'=g_3''=g_3$. All other common coupling constants $g_i$ ($i=1,...,6$) 
appear in exactly the same way once the $S_4$ and $A_4$ indices are expanded
out. As identical equations yield identical results, the first line of
Eq.~\eqref{a4-flavon} gives rise to an alignment which suffers from the same
ambiguity as the alignment of Eq.~\eqref{phiS-alignment-mod}:  the ratio of
the VEVs of the two doublet flavon components,
i.e. $\frac{\langle\eta_{\nu,1}\rangle}{\langle\eta_{\nu,2}\rangle}$, remains
a free parameter. This ambiguity can be removed by introducing the driving
field $\tilde\varphi_\nu^0$ in the ${\bf 3}$ representation of $S_4$. As
$\varphi_\nu$ furnishes a ${\bf 3'}$ representation, we only get one coupling,
proportional to $g_7$, at the renormalisable level. With $\varphi_\nu$ already being
aligned in the $(1,1,1)^T$ direction, the resulting $F$-term condition
enforces equal VEVs for $\eta_{\nu,1}$ and $\eta_{\nu,2}$. In this minimal and
elegant way we have obtained the structure of the $S_4$ alignments given
in Eq.~\eqref{s4-align-nu}, with
\be
v_\nu^2=-\frac{1}{3g_4} \left[g_5+g_6
  \left(\frac{g_2}{2g_3}\right)^2\right]u_\nu^2 \ ,\qquad
w_\nu=-\frac{g_2}{2g_3} u_\nu \ .
\ee
As in the AF model, the VEV $u_\nu$ is driven to a non-zero value due to $m^2_{\xi_\nu}<0$.

The remaining terms of the flavon superpotential of Eq.~\eqref{s4-flavon-nu}
are responsible for driving the VEV of the $\zeta_\nu$ flavon in the ${\bf
  1'}$ representation of $S_4$. As the driving field $D^0$ is completely neutral under the
imposed discrete symmetries, compare Table~\ref{tab-S4-2}, we additionally encounter
Higgs fields as well as flavons of the charged lepton sector. The former do
not play any role in aligning the flavons as the $S_4$ breaking is
typically assumed to occur around the GUT scale. Therefore we can safely set
$f_3=0$. This leaves us with an $F$-term condition that intertwines the
$\zeta_\nu$ flavon with the flavons $\eta_\mu$ and $\eta_e$ of the
charged lepton sector. A simple way to disentangle the flavons of the two
sectors was put forward in \cite{Antusch:2011sx}. The idea is to introduce a
second driving field with identical quantum numbers as $D^0$. Then one obtains
two $F$-term conditions which are identical in their structure but involve
independent coupling constants, $f_i$ for one driving field and $f_i'$ for the
other. As both conditions must be satisfied, one can linearly combine the
equations and thus find a unique solution for the VEV of the $\zeta_\nu$ flavon,
\be
z_\nu^2 = - \frac{f_4^{}f_2'-f_2^{}f_4'}{f_1^{}f_2'-f_2^{}f_1'} M_0^2  \ .\label{znuvev}
\ee

Turning to the flavon alignment of the charged lepton sector, the
renormalisable driving superpotential takes the form
\bea
W_{S_4}^{\mathrm{flavon},\ell}&=& 
 \varphi^0_\ell (h_1 \varphi_\ell \varphi_\ell + h_2 \varphi_\ell \eta_\mu) 
+\xi^0_\ell (h_3 \varphi_\ell \varphi_\ell + h_4 \eta_\mu \eta_\mu) 
\notag \\[2mm]
&& + \eta_\ell^0 (M_2 \eta_e + h_5\zeta_\nu \eta_e +h_6 \eta_\mu \eta_\mu
+h_7\varphi_\ell \varphi_\ell) \ . \label{s4-align-charged}
\eea
The alignments of the flavon fields $\varphi_\ell$ and $\eta_\mu$ are
constrained by the first line of Eq.~\eqref{s4-align-charged}. Writing
$\langle {\varphi_\ell}_i\rangle = t_i$ and $\langle {\eta_\mu}_i\rangle =
r_i$, the $F$-term conditions read 
\be
2h_1\begin{pmatrix} t_1^2-t_2t_3\\t_3^2-t_1t_2\\t_2^2-t_3t_1 \end{pmatrix} 
+
h_2 r_1 \begin{pmatrix} t_2\\t_3\\t_1 \end{pmatrix} 
+
h_2 r_2 \begin{pmatrix} t_3\\t_1\\t_2 \end{pmatrix} 
= \begin{pmatrix} 0\\0\\0 \end{pmatrix} ,\label{ftermS4}
\ee
\be
h_3 (t_1^2+2t_2t_3) +2 h_4 r_1 r_2 ~=~ 0 \ .\label{ftermS4-2}
\ee
The most general solution to these equations can be obtained by a
straightforward calculation.  As a result one finds two solutions,
\be
\langle \varphi_\ell \rangle = v_\ell 
\begin{pmatrix} 0\\1\\0 \end{pmatrix} \ , \qquad
\langle \eta_\mu \rangle =w_\mu 
\begin{pmatrix} 0\\1 \end{pmatrix} \ ,\qquad
v_\ell=-\frac{h_2}{2h_1}\,w_\mu \ ,\label{s4-align-char}
\ee
as well as
\be
\langle \varphi_\ell \rangle = v_\ell 
\begin{pmatrix} 1\\1\\1 \end{pmatrix} \ , \qquad
\langle \eta_\mu \rangle=w_\mu 
\begin{pmatrix} 1\\- 1 \end{pmatrix} \ ,\qquad
v_\ell^2=\frac{2h_4}{3h_3} \, w_\mu^2 
\ .
\ee
Clearly, alignments which are derived from these by application of an
arbitrary $S_4$ transformation also satisfy the $F$-term conditions of
Eqs.~(\ref{ftermS4},\ref{ftermS4-2}).\footnote{For a discussion of why such
$S_4$ symmetry transformed solutions lead to the same physics see \cite{Hagedorn:2010th}.}
In order to get rid of the second solution we can use the idea of 
adding a second $\xi^0_\ell$ driving field \cite{Antusch:2011sx}, thus effectively splitting the
$F$-term condition of Eq.~\eqref{ftermS4-2}  into two equations,
\be
 (t_1^2+2t_2t_3)= 0 \ ,\qquad   r_1 r_2 =0 \ .
\ee
Then the remaining solution, i.e. Eq.~\eqref{s4-align-char}, is just what we
have used in the discussion of the charged lepton flavour structure, see Eq.~\eqref{s4-align-char0}.

Having fixed the alignment of $\varphi_\ell$ and $\eta_\mu$ we can now proceed
with the alignment of the flavon doublet $\eta_e$.  This is determined from
the $F$-term condition of the driving field $\eta^0_\ell$ in
Eq.~\eqref{s4-align-charged}. Inserting the VEVs of $\varphi_\ell$,
$\eta_\mu$ and $\zeta_\nu$, and writing $\langle \eta_{e,i}\rangle = q_i$ we obtain 
\be
M_2 \begin{pmatrix} q_1 \\q_2 \end{pmatrix}
+ h_5 z_\nu \begin{pmatrix} q_1 \\-q_2 \end{pmatrix}
+ h_6 w_\mu^2 \begin{pmatrix} 1\\0\end{pmatrix}
+ h_7 v_\ell^2 \begin{pmatrix} 1\\0 \end{pmatrix} =
\begin{pmatrix} 0\\0 \end{pmatrix} .
\ee
As $z_\nu$ is already fixed, $q_2=0$ and we find the $\eta_e$ alignment of
Eq.~\eqref{s4-align-char0} with
\be
 w_e = -\frac{h_6 w_\mu^2+h_7 v_\ell^2}{M_2+h_5 z_\nu} \ .
\ee
Finally, we have to consider the terms of $W_{S_4}^{\mathrm{flavon},\nu}$
proportional to the driving field $D^0$, see
Eq.~\eqref{s4-flavon-nu}. Repeating the discussion above Eq.~\eqref{znuvev},
the $\eta_\mu$ and $\eta_e$ VEVs are simultaneously driven to non-vanishing
values with
\be
w_\mu w_e = - \frac{f_4^{}f_1'-f_1^{}f_4'}{f_2^{}f_1'-f_1^{}f_2'} M_0^2  \ .
\ee

We conclude this section by sketching how the hierarchy 
of charged lepton masses might be obtained in our setup. We will adopt a very rough
approximation, dropping all order one coefficients and keeping only orders of
magnitude. Everything must be explained in terms of the input mass parameters
$M_0$ and $M_2$ of Eqs.~(\ref{s4-flavon-nu},\ref{s4-align-charged}) as well as
the involved messenger masses. Assuming
\be
M_2  ~\sim~ 1000 \,M_0 \ ,
\ee
we get
\be
v_\ell \sim w_\mu \sim 10 \,M_0 \ , \qquad
w_e \sim \frac{1}{10}\,  M_0 \ .
\ee
This yields a hierarchy of about 100 between the muon and the electron
mass. Setting the messenger masses $M_{\Omega_i}$ at a scale of around
$100M_0$ fixes the ratio between the tau and the muon mass at a value of order
10. The tau mass itself is then set to be approximately~$\frac{v_d}{10}$.

%%%%%%%%%%%%%%%%%%%%%%%%%%%%%%%%%%%%%%%%%

%%%%%%%%%%%%%%%%%%%%%%%%%%%%%%%%%%%%%%%%%

%%%%%%%%%%%%%%%%%%%%%%%%%%%%%%%%%%%%%%%%%

\section{\label{sec-analytic}Trimaximal mixing in terms of perturbative deviations to tri-bimaximal mixing}
\cleqn

In this section 
we make a perturbative and analytic study of the deviations to TB mixing
which are predicted by the $A_4$ and $S_4$ models of TM mixing, subject to the
phenomenological constraint that the reactor angle is perturbatively small,
which enables TM mixing to be viewed as a perturbative expansion around the TB limit.
This is natural from the point of view of $S_4$ models where TB mixing arises at LO,
broken by higher order corrections which preserve TM mixing.
It also enables alternative phenomenological proposals to be viewed and compared on
the same footing. 

Our starting point is the right-handed neutrino mass matrix of the $A_4$
model in Eq.~\eqref{mR}. It can be written as the sum of a matrix that
preserves TB mixing and a matrix that violates it, 
\be
M_R= M_R^{\mathrm{TB}}+\Delta M_R \ ,\label{MR1+2}
\ee
where
\be
M_R^{\mathrm{TB}}~=~ \alpha \begin{pmatrix} 2&-1&-1\\-1&2&-1\\-1&-1&2\end{pmatrix}
+\beta  \begin{pmatrix}1&0&0\\0&0&1\\0&1&0 \end{pmatrix}
+\gamma \begin{pmatrix}0&1&1\\1&1&0\\1&0&1 \end{pmatrix}
\ ,\label{mR1}
\ee
and 
\be
\Delta M_R = \Delta \begin{pmatrix}
0&1&-1\\1&-1&0\\-1&0&1\end{pmatrix}
\ ,\label{mR2}
\ee
with
\be
\Delta = \mbox{$\frac{1}{2}$}(\gamma''-\gamma') \ , \qquad
\gamma=\mbox{$\frac{1}{2}$} (\gamma'+\gamma'') \ .
\ee
In the $S_4$ model the explicit form of $\Delta M_R $ is given in Eq.~\eqref{DeltaMR}.
In both the $A_4$ and $S_4$ models the TB violating matrix is required to be small,
\be 
|\Delta| \ll |\alpha|,|\beta| \ .\label{approx}
\ee
This assumption is necessary in order to meet the
experimental constraints that any deviations from TB mixing should be
small. The parameter $\gamma$ on the other hand need not be small as it does
not break the TB pattern. 

Since $M_R^{\mathrm{TB}}$ is diagonalised by the TB mixing matrix
$U_{\mathrm{TB}}$, this enables $M_R$ to be diagonalised perturbatively by
expanding about the TB mixing case.  
Writing the matrix $U_R$ which diagonalises the right-handed neutrino mass matrix
in terms of its column vectors $\Phi_i$,
\be
U_R ~=~ (\Phi_1 , \Phi_2 , \Phi_3) \ ,\label{urCOLUMN}
\ee 
we can expand $U_R$, for small deviations from TB mixing, in linear
approximation around its TB form using
\be
\Phi_i ~=~ \Phi_i^{\mathrm{TB}} + \Delta \Phi_i   \ ,
\qquad
 \Delta \Phi_i  
~=~  \sum_j \alpha_{ij} \Phi^{\mathrm{TB}}_j  \ ,
\ee
where $ \Phi^{\mathrm{TB}}_i$ are just the columns of the TB mixing matrix.
As shown in Appendix \ref{MRdiag},
due to the unitarity of $U_R$ and the special form of the mass matrix $M_R$ in Eq.~\eqref{MR1+2},
the only non-zero
parameter is  $\alpha_{13}=-\alpha_{31}^\ast$
whose dependence on the input parameters $\alpha,\beta,\gamma,\Delta$ is given in
Eqs.~(\ref{Re_alpha13},\ref{Im_alpha13}). The fact that only $\alpha_{13}=-\alpha_{31}^\ast$
is non-zero implies that $U_R$ is of TM form as expected. Furthermore, 
since,
\be
U_R^T M^{}_R U_R^{} ~=~ M_R^{\mathrm{diag} } \ ,\label{Mrdiag}
\ee
it is then straightforward to derive the lepton mixing matrix $U_{\mathrm{PMNS}}$,
as in Eq.~\eqref{URUnu}, 
\be
U_{\mathrm{PMNS}}~=~\frac{m_D}{y v_u} \, U_R  \ .\label{URUnu2}
\ee
Due to the trivial
structure of $m_D$ as well as a diagonal charged lepton sector, the PMNS mixing
matrix can thus be directly obtained from $U_R$ by  permuting the
second and the third row as well as multiplying the Majorana phase matrix  $P$
on the right and another phase matrix $P'$ on the left, leading to $U_{\mathrm{PMNS}}=U_{\mathrm{TM}}$
where,
\be
U_{\mathrm{TM}}~\approx~P'   \begin{pmatrix} 
\frac{2}{\sqrt{6}} &\frac{1}{\sqrt{3}} &-\frac{2}{\sqrt{6}} \alpha^\ast_{13}\\
-\frac{1}{\sqrt{6}}+  \frac{1}{\sqrt{2}} \alpha_{13}& \frac{1}{\sqrt{3}}&
\phantom{-}\frac{1}{\sqrt{2}}+\frac{1}{\sqrt{6}} \alpha^\ast_{13}\\
-\frac{1}{\sqrt{6}} -  \frac{1}{\sqrt{2}} \alpha_{13}&\frac{1}{\sqrt{3}}&
-\frac{1}{\sqrt{2}}+\frac{1}{\sqrt{6}} \alpha^\ast_{13} 
\end{pmatrix}  P\ .\label{PMNS-ex} 
\ee
The matrix $P'$ has to be chosen such that the PMNS matrix without Majorana
phases is brought to the standard PDG form where
the 2-3 and 3-3 elements are real and the mixing angles are all between
$0^\circ$ and $90^\circ$. In linear approximation, the required form of $P'$
becomes
\be
P'~\approx~\mathrm{diag}(1,a_+,- a_-) \ , \qquad
a_\pm\,=\,1\pm i \cdot \frac{\mathrm{Im}(\alpha_{13}) }{\sqrt{3}} \ .
\ee
Multiplying this explicit form of the phase matrix $P'$ we obtain a  mixing
matrix that is consistent with the standard PDG phase conventions. 

It is useful to compare the TM mixing matrix in 
Eq.~\eqref{PMNS-ex} to a general parametrisation of the 
PMNS mixing matrix in terms of deviations from TB mixing \cite{King:2007pr}, 
\begin{eqnarray}
U_{\mathrm{PMNS}} \approx
\left( \begin{array}{ccc}
{\frac{2}{\sqrt{6}}}(1-\frac{1}{2}s)  & \frac{1}{\sqrt{3}}(1+s) & \frac{1}{\sqrt{2}}re^{-i\delta } \\
-\frac{1}{\sqrt{6}}(1+s-a + re^{i\delta })  & \phantom{-}\frac{1}{\sqrt{3}}(1-\frac{1}{2}s-a- \frac{1}{2}re^{i\delta })
& \frac{1}{\sqrt{2}}(1+a) \\
\phantom{-}\frac{1}{\sqrt{6}}(1+s+a- re^{i\delta })  & -\frac{1}{\sqrt{3}}(1-\frac{1}{2}s+a+ \frac{1}{2}re^{i\delta })
 & \frac{1}{\sqrt{2}}(1-a)
\end{array}
\right)P\ ,~~
\label{PMNS1}
\end{eqnarray}
where the deviation parameters $s,a,r$ are defined as
\cite{King:2007pr}, 
\begin{eqnarray}
&&
 \sin \theta_{12} = \frac{1}{\sqrt{3}}(1+s)\ , \qquad
\sin \theta_{23}  =  \frac{1}{\sqrt{2}}(1+a)\  , \qquad 
 \sin \theta_{13} =   \frac{r}{\sqrt{2}}\ .
\label{rsa}
\end{eqnarray}
This comparison yields
\be
s\,\approx\,0 \ , \quad
a\,\approx\, \frac{\mathrm{Re}\,(\alpha_{13})}{\sqrt{3} } \ , \quad
r \cos \delta \,\approx\, -\frac{2}{\sqrt{3} } \,\mathrm{Re}\,( \alpha_{13})
\ , \quad
\delta\,\approx\,\mathrm{arg}\,(\alpha_{13}) + \pi \ ,
\ee
where $\delta$ is the CP violating oscillation phase of the lepton sector,
and $\alpha_{13}$ is proportional to $\Delta$ as shown in Eqs.~(\ref{Re_alpha13},\ref{Im_alpha13}).
 Independently of the value of $\alpha_{13}$ we confirm the TM sum rules given in \cite{King:2009qt},
\be
s\approx 0\ , \qquad  a \approx -\frac{1}{2}r \cos \delta \ .  \label{a4sumrule}
\ee
We emphasise that the sum rules in Eq.~\eqref{a4sumrule} hold for any model of trimaximal mixing,
since it is always the case that phenomenology requires that it has 
approximate tri-bimaximal form. 
The above perturbative form of TM mixing in terms of TB deviation parameters 
is useful when comparing TM mixing to other proposed forms of the
mixing matrix. For example, if the reactor angle is measured to be sizeable, but the solar and atmospheric angles remain close to their tri-bimaximal
values, i.e. the deviation parameters in Eq.~(\ref{rsa}) take the form
$s\approx a \approx 0$ but $r\neq 0$, then the mixing matrix takes the ``tri-bimaximal-reactor'' (TBR) form
\cite{King:2009qt}.
Such a mixing has recently been obtained in an $S_4$ setup
\cite{Morisi:2011pm}. Alternative proposals
\cite{Ge:2011ih,He:2011kn,Xing:2011at,Zhou:2011nu,Araki:2011wn,Haba:2011nv,Meloni:2011fx,Chao:2011sp,Zhang:2011aw,Chu:2011jg,Dev:2011gi,Toorop:2011jn,Antusch:2011qg}
that have been put forward to accommodate the T2K result could similarly be
compared using the deviation parameters $s,a,r$. With future neutrino
oscillation experiments being able to not only accurately measure the
reactor angle, parametrised here as $r$, but also the atmospheric and solar deviation
parameters $a,s$ and eventually the CP violating oscillation phase $\delta$,
it is clear that relating these deviation parameters via sum rules comprise
the next step in discriminating different models of lepton masses and mixings.

%%%%%%%%%%%%%%%%%%%%%%%%%%%%%%%%%%

%%%%%%%%%%%%%%%%%%%%%%%%%%%%%%%%%%

%%%%%%%%%%%%%%%%%%%%%%%%%%%%%%%%%%

\section{\label{sec-concl}Conclusions}
In the well known direct models of  tri-bimaximal (TB) mixing, based on  $A_4$ and $S_4$, 
the TB mixing is enforced by a Klein symmetry $Z_2^S\times Z_2^U$ in the neutrino sector,
together with a $Z_3^T$ symmetry in the charged lepton sector, where a common
basis corresponds to a diagonal charged lepton mass matrix.
It is also well known that TB mixing can emerge from either $S_4$, which contains the generators $S,T,U$, or 
$A_4$, which contains $S,T$. In the case of $A_4$ the $U$ symmetry emerges accidentally
as a result of the absence of flavons in the ${\bf 1'}$ or ${\bf 1''}$ representations of $A_4$.
Such models are called ``direct models'' since (some of) the group generators remain
unbroken in different sectors of the low energy effective theory.
Although this simple and appealing picture is apparently shattered by the  
T2K results, which indicate a sizeable reactor angle $\theta_{13}$,
simple alternative possibilities such as trimaximal (TM) mixing remain.

We have argued that, in the framework of direct $A_4$ models,
the T2K results motivate adding the ``missing'' flavons ${\bf 1'}$ or ${\bf 1''}$, 
whose VEVs break the accidental $U$ symmetry but preserve the $S$ symmetry
in the neutrino sector corresponding to a $Z_2^S$ subgroup of $A_4$,
leading to a TM pattern of mixing as in Eq.~\eqref{TM}. 
We have studied the vacuum alignment of such an extended 
Altarelli-Feruglio $A_4$ family symmetry model including
additional flavons in the ${\bf 1'}$ and ${\bf 1''}$ representations and show that it
leads to TM mixing in which the second column of the
lepton mixing matrix consists of the column vector  $(1,1,1)^T/\sqrt{3}$.
However there are drawbacks to this approach.
To begin with, the higher order corrections need to be kept under control otherwise they
can wash out the desirable TM structure, however this can in principle be done
by formulating the theory at the renormalisable level. Furthermore, this
approach within $A_4$ provides no explanation for why the reactor angle should
be smaller than the atmospheric and solar angles. 

In order to overcome these drawbacks, we have proposed a renormalisable
$S_4$ model of leptons in which the ${\bf 1'}$ and ${\bf 1''}$ flavons of
$A_4$ are unified into a doublet of $S_4$ which is spontaneously broken to $A_4$
by a ${\bf 1'}$ flavon of $S_4$. 
We have studied the vacuum alignment in the 
$S_4$ model and shown that it predicts {\em accurate} TM neutrino mixing 
due to a preserved $Z_2^S$ symmetry in the neutrino sector,
where $Z_2^S \subset Z_2^S\times Z_2^U \subset S_4$.
In the $S_4$ model, the $U$ generator corresponding to 
$Z_2^U $ is broken by the ${\bf 1'}$ flavon of $S_4$,
which however only enters the neutrino sector at higher order,
resulting in {\em approximate} TB mixing. 

Although the $A_4$ and $S_4$ models of leptons presented here involve 
diagonal charged lepton mass matrices, when the models are extended to include quarks,
for example in the framework of $SU(5)$ unification, we would expect the charged lepton sectors
(but not the neutrino sectors) of these models to be modified. 
This could introduce additional contributions to lepton mixing
from the charged lepton sector. Interestingly both the $A_4$ and $S_4$ models here preserve form dominance and
hence predict zero leptogenesis,  independently of charged lepton mixing,
up to renormalisation group corrections.

It should be clear that a sizeable reactor angle as indicated by T2K
does not rule out the family symmetry approach, on the contrary it provides
additional input which helps to refine the symmetry approach.
In particular we have explored an $S_4$ model in which 
the leading order Klein symmetry $ Z_2^S\times Z_2^U$ associated with TB mixing
is broken at higher order down to a $Z_2^S$ subgroup capable of enforcing 
a simple TM mixing pattern.
We emphasise that the smoking gun signature of TM mixing is accurate trimaximal
solar mixing $s\approx 0$, together with the sum rule 
$2a + r \cos \delta \approx 0$, 
which can be tested in forthcoming high precision
neutrino oscillation experiments.

%%%%%%%%%%%%%%%%%%%%%%%%%%%%%%%%%%%%%%%%%

%%%%%%%%%%%%%%%%%%%%%%%%%%%%%%%%%%%%%%%%%

%%%%%%%%%%%%%%%%%%%%%%%%%%%%%%%%%%%%%%%%%

\section*{Acknowledgements}

The authors acknowledge support from the STFC Rolling Grant
No. ST/G000557/1. The research was partially supported by the EU ITN grant
UNILHC 237920 (Unification in the LHC era). 

%%%%%%%%%%%%%%%%%%%%%%%%%%%%%%%%%%%%%%%%%

%%%%%%%%%%%%%%%%%%%%%%%%%%%%%%%%%%%%%%%%%

%%%%%%%%%%%%%%%%%%%%%%%%%%%%%%%%%%%%%%%%%

\section*{Appendix}

\begin{appendix}

\section{\label{a4-s4-CGs}${\bs{S_4}}$ and ${\bs{A_4}}$  Clebsch-Gordan
  coefficients}
\cleqn

Finite groups can be defined in terms of their generating elements which
satisfy certain product rules, the so called presentation. Starting from these rules it
is possible to work out the matrix representations of these elements up to
an ambiguity related to the choice of basis. In the literature there exist
different 'standard' choices which all have advantages but also
disadvantages. As the Clebsch-Gordan coefficients depend on the chosen
basis, it is necessary to carefully define the basis in which a model is
constructed. 

The generators and Clebsch-Gordan coefficients of $S_4=\Delta(24)$ and
$A_4=\Delta(12)$ in a  basis where the triplets are explicitly real were
derived in a general way in \cite{Escobar:2008vc,Luhn:2007uq}. As was argued in
\cite{Altarelli:2005yx}, there exists a more suitable triplet basis for $A_4$
models in which
the order-three generator $T$ is brought to a diagonal and complex form. By
now this choice has become the standard or physical basis for direct models
\cite{King:2009ap}. The corresponding basis in the case of $S_4$ can  be
found for instance in \cite{Bazzocchi:2009pv}. As $A_4$ is a subgroup of $S_4$ it is
natural to express this relation also in terms of the generators where
the elements $S,T,U$ generate $S_4$, while $A_4$ is obtained by simply dropping
the $U$ generator \cite{King:2009mk}. 

Table~\ref{generatorss4a4} lists the generators of $S_4$ and $A_4$ in the
physical basis. The primed representations of $S_4$ differ only in the sign of
the $U$ generator. Dropping the $U$ generator we obtain $A_4$. It is clear
from the table that the doublet of $S_4$ becomes a reducible representation
under $A_4$, denoted by ${\bf 1''}$ and ${\bf 1'}$.
\begin{table}
$$
\begin{array}{c|ccc|c}\toprule
S_4 & A_4 & S & T & U \\\midrule
{\bf 1,1'} & {\bf 1} & 1&1 &\pm 1 \\[2mm]
{\bf 2} & \begin{pmatrix}{\bf 1''}\\{\bf 1'}\end{pmatrix} & 
\begin{pmatrix} 1 & 0 \\ 0&1 \end{pmatrix} 
&\begin{pmatrix} \omega & 0 \\ 0&\omega^2 \end{pmatrix}  
& \begin{pmatrix} 0& 1 \\ 1&0 \end{pmatrix} \\[4mm]
{\bf 3,3'} & {\bf 3}& 
\frac{1}{3}\begin{pmatrix} -1 & 2&2 \\ 2&-1&2 \\2&2&-1 \end{pmatrix} 
&\begin{pmatrix} 1&0&0\\ 0&\omega^2 &0 \\ 0&0&\omega \end{pmatrix}  
& \mp \begin{pmatrix} 1&0&0\\0&0& 1 \\ 0&1&0 \end{pmatrix} 
\\\bottomrule
\end{array}
$$
\caption{\label{generatorss4a4}The generators $S,T,U$ of $S_4$ and $S,T$ of
  $A_4$ as used in this article.} 
\end{table}

The $S_4$ product rules in the chosen basis are listed below, where we use 
the number of primes within the expression
\be
{\bs{\alpha}}^{(\prime)}  \otimes {\bs {\beta}}^{(\prime)} ~\rightarrow
~{\bs{\gamma}}^{(\prime)} \ , \label{CGnotation} 
\ee
to classify the results. We denote this number by $n$, e.g. in ${\bf 3}\otimes
{\bf 3}^\prime \rightarrow {\bf 3}^\prime$ we get $n=2$.

%%%%%%%%%%%%%%%%%%%%%%%%%%%
%%%%    S4 Clebsch Gordan coefficients    %%%%  BEGINNING
%%%%%%%%%%%%%%%%%%%%%%%%%%%

$$
\begin{array}{lll}
{\bf 1}^{(\prime)} \otimes {\bf 1}^{(\prime)} ~\rightarrow ~{\bf
  1}^{(\prime)} ~~
\left\{ \begin{array}{c} 
~\\n=\mathrm{even}\\~
\end{array}\right.
&%
\left.
\begin{array}{c} 
{\bf 1}^{\phantom{\prime}} \otimes {\bf 1}^{\phantom{\prime}} ~\rightarrow ~{\bf 1}^{\phantom{\prime}}\\
{\bf 1}^{{\prime}} \otimes {\bf 1}^{{\prime}} ~\rightarrow ~{\bf 1}^{\phantom{\prime}}\\
{\bf 1}^{\phantom{\prime}} \otimes {\bf 1}^{{\prime}} ~\rightarrow ~{\bf 1}^{{\prime}}
\end{array}
\right\}
&
\alpha \beta \ ,
\\[10mm]
{\bf 1}^{(\prime)} \otimes \;{\bf 2} \;~\rightarrow \;~{\bf 2}^{\phantom{(\prime)}}~~ \left\{
\begin{array}{c}
n=\mathrm{even} \\
n=\mathrm{odd}
\end{array} \right.
&%
\left.
\begin{array}{c}
{\bf 1}^{\phantom{\prime}} \otimes {\bf 2} ~\rightarrow ~{\bf 2} \\
{\bf 1}^{\prime} \otimes {\bf 2} ~\rightarrow ~{\bf 2}\\
\end{array}\;~
\right\}
&
 \alpha \begin{pmatrix} \beta_1 \\ (-1)^n \beta_2\end{pmatrix}  ,
\\[7mm]
{\bf 1}^{(\prime)} \otimes {\bf 3}^{(\prime)} ~\rightarrow ~{\bf 3}^{(\prime)}
~~ \left\{ \begin{array}{c}
~\\[3mm]n=\mathrm{even} \\[3mm]~
\end{array}\right.
&%
\left. 
\begin{array}{c}
{\bf 1}^{\phantom{\prime}} \otimes {\bf 3}^{\phantom{\prime}} ~\rightarrow ~{\bf 3}^{\phantom{\prime}}
\\
{\bf 1}^{{\prime}} \otimes {\bf 3}^{{\prime}} ~\rightarrow ~{\bf 3}^{\phantom{\prime}}
\\
{\bf 1}^{\phantom{\prime}} \otimes {\bf 3}^{{\prime}} ~\rightarrow ~{\bf 3}^{{\prime}}
\\
{\bf 1}^{{\prime}} \otimes {\bf 3}^{\phantom{\prime}} ~\rightarrow ~{\bf 3}^{{\prime}}
\end{array}
\right\}
&
 \alpha   \begin{pmatrix} \beta_1 \\  \beta_2\\\beta_3 \end{pmatrix}  ,
\\[12.2mm]
{\bf 2} \;\; \otimes \;\;{\bf 2} \;~\rightarrow \;~{\bf 1}^{(\prime)} ~~ \left\{\begin{array}{c}
n=\mathrm{even}\\
n=\mathrm{odd}
\end{array}\right.
&%
\left.
\begin{array}{c}
{\bf 2} \otimes {\bf 2} ~\rightarrow ~{\bf 1}^{\phantom{\prime}} \\
{\bf 2} \otimes {\bf 2} ~\rightarrow ~{\bf 1}^{{\prime}} 
\end{array}~\;
\right\}
&
 \alpha_1 \beta_2 + (-1)^n \alpha_2 \beta_1 \ , 
%\\[7mm]
%
%
%
%
%
\end{array}
$$
$$
\begin{array}{lll}
{\bf 2} \;\;\otimes \;\; {\bf 2} ~\;\rightarrow \;~{\bf 2}^{\phantom{(\prime)}} ~~ \left\{ \begin{array}{c}
~\\[-3mm] n=\mathrm{even}\\[-3mm]~
\end{array}\right.
&%
\left.
\begin{array}{c}
~\\[-3mm]
{\bf 2} \otimes {\bf 2} ~\rightarrow ~{\bf 2} \\[-3mm]~
\end{array}~~\,
\right\}
&
   \begin{pmatrix} \alpha_2 \beta_2 \\  \alpha_1\beta_1 \end{pmatrix} , 
\\[6mm]
{\bf 2}\;\; \otimes \; {\bf 3}^{{(\prime)}} ~\rightarrow ~{\bf 3}^{{(\prime)}} ~~ \left\{\begin{array}{c}
~\\[-2mm] n=\mathrm{even}\\ \\[2mm]
n=\mathrm{odd}\\[-2mm]~
\end{array}\right.
&%
\left.
\begin{array}{c}
{\bf 2} \otimes {\bf 3}^{\phantom{\prime}} ~\rightarrow ~{\bf 3}^{\phantom{\prime}} \\
{\bf 2} \otimes {\bf 3}^{{\prime}} ~\rightarrow ~{\bf 3}^{{\prime}} \\[3mm]
{\bf 2} \otimes {\bf 3}^{\phantom{\prime}} ~\rightarrow ~{\bf 3}^{{\prime}} \\
{\bf 2} \otimes {\bf 3}^{{\prime}} ~\rightarrow ~{\bf 3}^{\phantom{\prime}} 
\end{array}\;
\right\}
&
 \alpha_1 \begin{pmatrix} \beta_2 \\  \beta_3\\\beta_1 \end{pmatrix} + (-1)^n
\alpha_2 \begin{pmatrix} \beta_3 \\  \beta_1\\\beta_2 \end{pmatrix}  ,
\\[13.5mm]
{\bf 3}^{(\prime)} \otimes {\bf 3}^{(\prime)} ~\rightarrow ~{\bf 1}^{(\prime)}
~~ \left\{ \begin{array}{c}
~\\n=\mathrm{even}\\~
\end{array}\right.
&%
\left.\begin{array}{c}
{\bf 3}^{\phantom{\prime}} \otimes {\bf 3}^{\phantom{\prime}} ~\rightarrow ~{\bf 1}^{\phantom{\prime}}
\\
{\bf 3}^{{\prime}} \otimes {\bf 3}^{{\prime}} ~\rightarrow ~{\bf 1}^{\phantom{\prime}}
\\
{\bf 3}^{\phantom{\prime}} \otimes {\bf 3}^{{\prime}} ~\rightarrow ~{\bf 1}^{{\prime}}
\end{array}\right\}
&
 \alpha_1 \beta_1 +\alpha_2\beta_3+\alpha_3\beta_2 \ ,
\\[9mm]
{\bf 3}^{(\prime)} \otimes {\bf 3}^{(\prime)} ~\rightarrow ~{\bf 2}^{\phantom{(\prime)}} ~~
\left\{ \begin{array}{c}
~\\[-3mm]
n=\mathrm{even}\\ \\[1mm]
n=\mathrm{odd}\\[-4.5mm]~
\end{array}\right.
&%
\left.\begin{array}{c}
{\bf 3}^{\phantom{\prime}} \otimes {\bf 3}^{\phantom{\prime}} ~\rightarrow ~{\bf 2} \\
{\bf 3}^{{\prime}} \otimes {\bf 3}^{{\prime}} ~\rightarrow ~{\bf 2} \\[3mm]
{\bf 3}^{\phantom{\prime}} \otimes {\bf 3}^{{\prime}} ~\rightarrow ~{\bf 2} \\
\end{array}\;
\right\}
&
\begin{pmatrix} \alpha_2 \beta_2 +\alpha_3 \beta_1+\alpha_1\beta_3\\ 
(-1)^n(\alpha_3 \beta_3+\alpha_1\beta_2+\alpha_2\beta_1) \end{pmatrix} ,
\\[10.5mm]
{\bf 3}^{(\prime)} \otimes {\bf 3}^{(\prime)} ~\rightarrow ~{\bf 3}^{(\prime)}
~~ \left\{\begin{array}{c}
~\\n=\mathrm{odd}\\~
\end{array}\right.
&%
\left.\begin{array}{c}
{\bf 3}^{\phantom{\prime}} \otimes {\bf 3}^{\phantom{\prime}} ~\rightarrow ~{\bf 3}^{{\prime}}
\\
{\bf 3}^{\phantom{\prime}} \otimes {\bf 3}^{{\prime}} ~\rightarrow ~{\bf 3}^{\phantom{\prime}}
\\
{\bf 3}^{{\prime}} \otimes {\bf 3}^{{\prime}} ~\rightarrow ~{\bf 3}^{{\prime}}
\end{array}\right\}
& 
\begin{pmatrix} 
2 \alpha_1 \beta_1-\alpha_2\beta_3-\alpha_3\beta_2 \\  
2 \alpha_3 \beta_3-\alpha_1\beta_2-\alpha_2\beta_1 \\  
2 \alpha_2 \beta_2-\alpha_3\beta_1-\alpha_1\beta_3 
 \end{pmatrix} ,
\\[9mm]
{\bf 3}^{(\prime)} \otimes {\bf 3}^{(\prime)} ~\rightarrow ~{\bf 3}^{(\prime)}~~\left\{\begin{array}{c}
~\\n=\mathrm{even}\\~
\end{array}\right.
&%
\left.\begin{array}{c}
{\bf 3}^{\phantom{\prime}} \otimes {\bf 3}^{\phantom{\prime}} ~\rightarrow ~{\bf 3}^{\phantom{\prime}}
\\
{\bf 3}^{{\prime}} \otimes {\bf 3}^{{\prime}} ~\rightarrow ~{\bf 3}^{\phantom{\prime}}
\\
{\bf 3}^{\phantom{\prime}} \otimes {\bf 3}^{{\prime}} ~\rightarrow ~{\bf 3}^{{\prime}}
\end{array}\right\}
&
\begin{pmatrix} 
\alpha_2\beta_3-\alpha_3\beta_2 \\  
\alpha_1\beta_2-\alpha_2\beta_1 \\  
\alpha_3\beta_1-\alpha_1\beta_3 
 \end{pmatrix}  .
\end{array}\\[3mm]
$$

%%%%%%%%%%%%%%%%%%%%%%%%%%%
%%%%    S4 Clebsch Gordan coefficients    %%%%  END
%%%%%%%%%%%%%%%%%%%%%%%%%%%

The $A_4$ Clebsch-Gordan coefficients can be obtained from the above
expressions by simply dropping all primes and identifying the components of
the $S_4$ doublet ${\bf 2}$ as the ${\bf 1''}$ and ${\bf 1'}$ representations
of $A_4$, see Table~\ref{generatorss4a4}.
We thus find for the non-trivial products

%%%%%%%%%%%%%%%%%%%%%%%%%%%
%%%%    A4 Clebsch Gordan coefficients    %%%%  BEGINNING
%%%%%%%%%%%%%%%%%%%%%%%%%%%

$$
\begin{array}{lcl}
{\bf 1'} \otimes {\bf 1''} ~\rightarrow ~{\bf 1} 
&&
\alpha \beta \ ,\\[3mm]
{\bf 1'} \otimes {\bf 3} ~\rightarrow ~{\bf 3}  
&&
\alpha \begin{pmatrix} 
\beta_3 \\  
\beta_1 \\  
\beta_2 
 \end{pmatrix}  , \\[8mm]
{\bf 1''} \otimes {\bf 3} ~\rightarrow ~{\bf 3}  
&&
\alpha \begin{pmatrix} 
\beta_2 \\  
\beta_3 \\  
\beta_1 
 \end{pmatrix}  , \\[8mm]
{\bf 3} \otimes {\bf 3} ~\rightarrow ~{\bf 1} 
&&
\alpha_1\beta_1 +\alpha_2\beta_3+\alpha_3\beta_2 \ ,\\[2mm]
{\bf 3} \otimes {\bf 3} ~\rightarrow ~{\bf 1'} 
&&
\alpha_3\beta_3 +\alpha_1\beta_2+\alpha_2\beta_1 \ ,\\[2mm]
{\bf 3} \otimes {\bf 3} ~\rightarrow ~{\bf 1''} 
&&
\alpha_2\beta_2 +\alpha_3\beta_1+\alpha_1\beta_3 \ ,\\[3mm]
{\bf 3} \otimes {\bf 3} ~\rightarrow ~{\bf 3}  + {\bf 3} 
&&
\begin{pmatrix} 
2 \alpha_1 \beta_1-\alpha_2\beta_3-\alpha_3\beta_2 \\  
2 \alpha_3 \beta_3-\alpha_1\beta_2-\alpha_2\beta_1 \\  
2 \alpha_2 \beta_2-\alpha_3\beta_1-\alpha_1\beta_3 
 \end{pmatrix}
+
\begin{pmatrix} 
\alpha_2\beta_3-\alpha_3\beta_2 \\  
\alpha_1\beta_2-\alpha_2\beta_1 \\  
\alpha_3\beta_1-\alpha_1\beta_3 
 \end{pmatrix}   . 
\end{array}
$$

%%%%%%%%%%%%%%%%%%%%%%%%%%%
%%%%    A4 Clebsch Gordan coefficients    %%%%  END
%%%%%%%%%%%%%%%%%%%%%%%%%%%

\section{\label{MRdiag}Perturbative diagonalisation of $\bs{M_R}$}
\cleqn

In this Appendix we present a general method for a perturbative diagonalisation
of the right-handed neutrino mass matrix and the determination
of the PMNS matrix in the diagonal charged lepton mass basis.
The method is analogous to the perturbative expansion of the effective neutrino mass
matrix developed in \cite{King:2010bk}, which is mainly applicable to the case of a diagonal
right-handed neutrino mass matrix as encountered in 
indirect models. We show that, in the case of the direct models in this paper, 
it is sufficient to consider the matrix $U_R$ which perturbatively diagonalises
the right-handed neutrino mass matrix, since the PMNS matrix is then
simply obtained from $U_R$ due to the trivial structure of the Dirac mass matrix.

Writing the matrix $U_R$ as in Eq.~\eqref{urCOLUMN}, 
$U_R = (\Phi_1 , \Phi_2 , \Phi_3)$,  
we can express the mass matrix $M_R$, see Eq.~\eqref{Mrdiag}, as
\be
M_R ~=~ U_R^\ast M_R^{\mathrm{diag}} \, U_R^{\dagger} ~=~ \sum_i M_i \, \Phi^{\ast}_i \Phi^\dagger_i \ ,\label{MR}
\ee 
where $M_i$ are the eigenvalues of $M_R$.
For small deviations from TB mixing, we can expand this equation in linear
approximation around its TB form using
\be
M_R^{} = M_R^{\mathrm{TB}} + \Delta M^{}_R \ , \qquad 
M_i ~=~ M_i^{\mathrm{TB}} + \Delta M^{}_i  \ , \qquad
\Phi_i ~=~ \Phi_i^{\mathrm{TB}} + \Delta \Phi_i   \ .
\ee
Here $M_R^{\mathrm{TB}}$ is the part of $M_R$ that does not depend on the
small parameter $\Delta$. Its eigenvalues are
\be
M_1^{\mathrm{TB}}= 3\alpha+\beta-\gamma \ , \qquad
M_2^{\mathrm{TB}}=\beta+2\gamma\ , \qquad
M_3^{\mathrm{TB}} =3\alpha-\beta+\gamma\ ,
\ee
with eigenvectors
\be
\Phi^{\mathrm{TB}}_1 = \frac{1}{\sqrt{6}}\begin{pmatrix}2\\-1\\-1\end{pmatrix} , \qquad
\Phi^{\mathrm{TB}}_2 = \frac{1}{\sqrt{3}}\begin{pmatrix}1\\1\\1\end{pmatrix} , \qquad
\Phi^{\mathrm{TB}}_3 = \frac{1}{\sqrt{2}}\begin{pmatrix}0\\-1\\1\end{pmatrix}.
\ee
Following \cite{King:2010bk}, we parameterise $\Delta\Phi_i$ in the basis of
these TB column vectors $\Phi_i^{\mathrm{TB}}$,
\be
 \Delta \Phi_i  
~=~  \sum_j \alpha_{ij} \Phi^{\mathrm{TB}}_j  \ .
\ee
The parameters $\alpha_{ij}$, which are a measure of how much the mixing
deviates from the TB pattern, are small. In order to determine their
dependence on the input parameters $\alpha,\beta,\gamma,\Delta$, we first
observe that $U_R$ must be unitary, i.e.  
\be
\delta_{ij}  ~=~
\left( \Phi_i^{\mathrm{TB}} + \sum_k \alpha_{ik} \Phi^{\mathrm{TB}}_k  \right)^\dagger
\left( \Phi_j^{\mathrm{TB}} + \sum_l \alpha_{jl} \Phi^{\mathrm{TB}}_l
\right) ~\approx~ \delta_{ij} +\alpha_{ij}^\ast + \alpha_{ji} \ ,
\ee
where we have dropped second order terms. As a consequence of 
unitarity we thus get
\be
\alpha_{ji} ~\approx~ - \alpha_{ij}^\ast \ . \label{unitarity-cond}
\ee
Expanding Eq.~\eqref{MR} around its TB structure and keeping only terms linear in
small deviations, we get
\bea
\Delta M_R^{}  &=& M_R^{} - M_R^{\mathrm{TB}}  \notag\\[2mm]
&\approx & \sum_i \left( \Delta M_i^{} \, \Phi^{\mathrm{TB}}_i {\Phi^{\mathrm{TB}}_i}^T  
+ M^{\mathrm{TB}}_i \,  \Delta \Phi^{\ast}_i {\Phi^{\mathrm{TB}}_i}^T 
+ M^{\mathrm{TB}}_i \,   \Phi^{\mathrm{TB}}_i {\Delta \Phi^{}_i}^\dagger 
\right) \label {deltaM-cond}\\
&\approx& \sum_i \left( \Delta M^{}_i \, \Phi^{\mathrm{TB}}_i {\Phi^{\mathrm{TB}}_i}^T 
+ M^{\mathrm{TB}}_i \,  \sum_j \alpha^\ast_{ij} \Phi^{\mathrm{TB}}_j {\Phi^{\mathrm{TB}}_i}^T 
+ M^{\mathrm{TB}}_i \,  \sum_j  \Phi^{\mathrm{TB}}_i {\alpha^\ast_{ij} \Phi^{\mathrm{TB}}_j}^T 
\right)\ . \notag
\eea
With $\Delta M_R$ corresponding to the TB breaking contribution to $M_R$, as in Eq.~\eqref{DeltaMR},
\be
\Delta M_R~=~\Delta \begin{pmatrix}
0&1&-1\\1&-1&0\\-1&0&1 
\end{pmatrix} ,
\ee
it is now possible to calculate the parameters $\alpha_{ij}$ and $\Delta M_i$
from Eq.~(\ref{deltaM-cond}). 
Counting the number of unknowns we get 6 real parameters for $\Delta M_i$ as
well as $3+6$ real parameters for $\alpha_{ij}$ where we have accounted for
the unitarity constraint of Eq.~(\ref{unitarity-cond}).\footnote{Note that
  $\alpha_{ii}$ must be purely imaginary.}
On the other hand Eq.~(\ref{deltaM-cond}) yields 6 complex conditions, 3 from
the diagonal entries and 3 from the off-diagonals ($\Delta M_R$ is
symmetric). As the number of unknowns is bigger than the number of conditions
we do not find a unique solution. Note however that the effect of
$\alpha_{ii}$ can always be absorbed in $\Delta M_i$. Hence we can remove 3
unknowns by setting $\alpha_{ii}=0$ without loss of generality.

The remaining 12 unknowns can be determined by sandwiching
Eq.~(\ref{deltaM-cond}) between ${\Phi^{\mathrm{TB}}_k}^T$ and
${\Phi^{\mathrm{TB}}_l}^{}$,
\be
{\Phi^{\mathrm{TB}}_k}^T \Delta M_R \, {\Phi^{\mathrm{TB}}_l}^{} ~\approx~ 
\left( \Delta M_k \delta_{kl}  
+ M^{\mathrm{TB}}_l \, \alpha_{lk}^\ast 
+ M^{\mathrm{TB}}_k \, \alpha_{kl}^\ast 
\right)\ .
\ee
Explicit calculation of the left-hand side gives
\bea
0 &\approx& M^{\mathrm{TB}}_1 \alpha_{12}^\ast - M^{\mathrm{TB}}_2 \alpha_{12} \ , \label{cond12}\\
0 &\approx& M^{\mathrm{TB}}_2 \alpha_{23}^\ast - M^{\mathrm{TB}}_3 \alpha_{23}  \ ,  \label{cond23} \\
-\sqrt{3}\, \Delta &\approx& M^{\mathrm{TB}}_1
\alpha_{13}^\ast  - M^{\mathrm{TB}}_3 \alpha_{13} \ , \label{cond13} \\[2mm]
0&\approx&\Delta M_1 \approx \Delta M_2 \approx \Delta M_3  \label{cond33} \ .
\eea
Notice that the TB breaking parameter $\Delta$ does not give rise to
corrections to the mass eigenvalues in linear approximation. 
Eqs.~(\ref{cond12}) and (\ref{cond23}) tell us that $\alpha_{12}=\alpha_{23}=0$
unless $|M^{\mathrm{TB}}_1| = |M^{\mathrm{TB}}_2|$ and $|M^{\mathrm{TB}}_2| =
|M^{\mathrm{TB}}_3|$, respectively. 
Eq.~(\ref{cond13}) can be used to determine the complex valued $\alpha_{13}$
in terms of the parameters $\alpha$, $\beta$, $\gamma$ and $\Delta$ of the
right-handed neutrino mass matrix $M_R$. 
A straightforward but tedious calculation yields
\bea
\mathrm{Re}\left(\alpha_{13} \right)&\approx& -\,\frac{\sqrt{3}}{2} \cdot
 \left[
\mathrm{Re}\left( \frac{\Delta}{\beta-\gamma} \right)  \,
+ 
\,\mathrm{Im} \left( \frac{\Delta}{\beta-\gamma} \right) 
\frac{\mathrm{Im} \left(
  \frac{3\alpha}{\beta-\gamma} \right)  }{\mathrm{Re} \left(
  \frac{3\alpha}{\beta-\gamma} \right)} 
\right]\ , \label{Re_alpha13}\\[2mm]
\mathrm{Im}\left(\alpha_{13} \right) &\approx& \frac{\sqrt{3}}{2} \cdot \frac{\mathrm{Im} \left( \frac{\Delta}{\beta-\gamma} \right) }{\mathrm{Re} \left(
  \frac{3\alpha}{\beta-\gamma} \right)} \ .
 \label{Im_alpha13}
 \eea
Notice that $\alpha_{13}$ is proportional to the TB violating parameter $\Delta$, 
which, in the $S_4$ model, is proportional to the
$\zeta_\nu$ flavon VEV as shown in Eq.~\eqref{inputs}.
Once $\alpha_{13}$ is determined, the matrix $U_R$, and hence the PMNS matrix,
are fully determined, as discussed in Section~\ref{sec-analytic}.

\end{appendix}

%%%%%%%%%%%%%%%%%%%%%%%%%%%%%%%%%%%%%%%%%

%%%%%%%%%%%%%%%%%%%%%%%%%%%%%%%%%%%%%%%%%

%%%%%%%%%%%%%%%%%%%%%%%%%%%%%%%%%%%%%%%%%

\end{document}